\documentclass[aps,pre,twocolumn,showpacs,groupedaddress,
               showkeys,amsmath,amssymb]{revtex4}

\newif\ifpdf
\ifx\pdfoutput\undefined
\pdffalse 
\else
\pdfoutput=1
\pdftrue 
\fi

\ifpdf 
\usepackage[pdftex]{graphicx} 
\else
\usepackage[dvips]{graphicx} 
\fi

\usepackage{graphicx}
\usepackage{bm}
\usepackage{bbm}
\usepackage{txfonts}

\usepackage[applemac]{inputenc}
\usepackage[T1]{fontenc}  
\usepackage{ae}          

\usepackage{epic}
\usepackage{eepic}
\usepackage{pifont}

\usepackage{nicefrac}
\newcommand{\textcomm}[1]{}

\begin{document}

\title{Comparison of phase-field models for surface diffusion}

\author{Clemens M\"uller-Gugenberger}
\affiliation{ Institut f\"ur Festk\"orperforschung,  Forschungszentrum J\"ulich,
D-52425 J\"ulich, Germany}

\author{Robert Spatschek} \affiliation{Center for Interdisciplinary Research on Complex Systems,
Northeastern University, Boston, MA 02115, USA}
\affiliation{Institut f\"ur Festk\"orperforschung,  Forschungszentrum J\"ulich,
D-52425 J\"ulich, Germany}

\author{Klaus Kassner} \affiliation{Institut f\"ur Theoretische
  Physik, Otto-von-Guericke-Universit\"at Magdeburg,
  Postfach 4120, D-39016 Magdeburg, Germany}
\date{10 November 2007}

\begin{abstract}
  The description of surface-diffusion controlled dynamics via the
  phase-field method is less trivial than it appears at first sight.
  A seemingly straightforward approach from the literature is shown to
  fail to produce the correct asymptotics, albeit in a subtle manner.
  Two models are constructed that approximate known sharp-interface
  equations without adding undesired constraints. Linear stability of
  a planar interface is investigated for the resulting phase-field
  equations and shown to reduce to the desired limit. Finally,
  numerical simulations of the standard and a more sophisticated model
  from the literature as well as of our two new models are performed
  to assess the relative merits of each approach. The results suggest
  superior performance of the new models in at least some situations.

\end{abstract}


\maketitle


\newcommand{\s}{{s}}
\newcommand{\uc}{q}
\newcommand{\vel}{\mathbf{v}}
\newcommand{\Vel}{{\mathbf V}}
\newcommand{\gprime}{g^{\prime}}
\newcommand{\abs}[1]{\vert#1\vert}
\newcommand{\normhat}{\hat{{\mathbf n}}}
\newcommand{\tanghat}{\hat{{\mathbf t}}}
\newcommand{\tangahat}{{\hat{\mathbf t}_1}}
\newcommand{\tangbhat}{{\hat{\mathbf t}_2}}
\newcommand{\chat}{\hat{c}}
\newcommand{\ghat}{\hat{G}}
\newcommand{\ez}{{\mathbf e}_z}
\newcommand{\ex}{{\mathbf e}_x}
\newcommand{\er}{{\mathbf e}_r}
\newcommand{\es}{{\mathbf e}_s}
\newcommand{\eu}{{\mathbf e}_u}
\newcommand{\norm}{{\mathbf n}}
\newcommand{\tang}{{\mathbf t}}
\newcommand{\tanga}{{\mathbf t}_1}
\newcommand{\tangb}{{\mathbf t}_2}
\newcommand{\rvec}{{\mathbf r}}
\newcommand{\Rvec}{{\mathbf R}}
\newcommand{\Uvec}{{\mathbf U}}
\newcommand{\wel}{{\mathbf v}}
\newcommand{\Wel}{{\mathbf V}}
\newcommand{\Avec}{{\mathbf A}}
\newcommand{\orderof}[1]{{\mathcal O}(#1)}
\newcommand{\linop}{{\mathcal L}}
\newcommand{\nablatwod}{{\nabla_{\rm 2D}}}
\newcommand{\lapltwod}{{\Delta_{\rm 2D}}}
\newcommand{\Er}{\mathcal{E}_r}
\newcommand{\Es}{{\mathcal E}_\s}
\newcommand{\Eu}{{\mathcal E}_\uc}
\newcommand{\Ei}{{\mathcal E}_i}
\newcommand{\Ej}{{\mathcal E}_j}
\newcommand{\Ealph}{{\mathcal E}_\alpha}
\newcommand{\Ebet}{{\mathcal E}_\beta}
\newcommand{\Ehr}{{\mathcal E}^r}
\newcommand{\Ehs}{{\mathcal E}^\s}
\newcommand{\Ehu}{{\mathcal E}^\uc}
\newcommand{\Ehi}{{\mathcal E}^i}
\newcommand{\Ehj}{{\mathcal E}^j}
\newcommand{\Ehalph}{{\mathcal E}^\alpha}
\newcommand{\Ehbet}{{\mathcal E}^\beta}
\newcommand{\Ehgam}{{\mathcal E}^\gamma}
\newcommand{\Ehmu}{{\mathcal E}^\mu}
\newcommand{\Ehnu}{{\mathcal E}^\nu}

\newcommand{\balpha}{{\bar{\alpha}}}
\newcommand{\bbeta}{{\bar{\beta}}}
\newcommand{\bgam}{{\bar{\gamma}}}
\newcommand{\bmu}{{\bar{\mu}}}
\newcommand{\bnu}{{\bar{\nu}}}

\newcommand{\Ehbalph}{{\mathcal E}^{\balpha}}
\newcommand{\Ehbbet}{{\mathcal E}^{\bbeta}}
\newcommand{\Ehbgam}{{\mathcal E}^{\bgam}}

\newcommand{\ssigma}{{\tilde{\sigma}}}
\newcommand{\Ssigma}{{\tilde{\Sigma}}}
\newcommand{\sigss}{\sigma_{\s\s}}
\newcommand{\siguu}{\sigma_{\uc\uc}}
\newcommand{\sigrr}{\sigma_{rr}}
\newcommand{\signn}{\sigma_{nn}}
\newcommand{\sigbalphbbet}{\sigma_{\balpha\bbeta}}
\newcommand{\sigbalphbalph}{\sigma_{\balpha\balpha}}

\newcommand{\phinul}{\phi^{(0)}}
\newcommand{\phione}{\phi^{(1)}}
\newcommand{\phitwo}{\phi^{(2)}}
\newcommand{\Phinul}{\Phi^{(0)}}
\newcommand{\Phione}{\Phi^{(1)}}
\newcommand{\Phitwo}{\Phi^{(2)}}

\newcommand{\phizero}{\phi_{0}}
\newcommand{\Phizero}{\Phi_{0}}

\newcommand{\phikalph}{\phi,_\alpha}
\newcommand{\phikbet}{\phi,_\beta}
\newcommand{\phikmu}{\phi,_\mu}
\newcommand{\Phikalph}{\Phi,_\alpha}
\newcommand{\Phikbet}{\Phi,_\beta}
\newcommand{\Phikmu}{\Phi,_\mu}
\newcommand{\Phikrho}{\Phi,_\rho}
\newcommand{\Phiks}{\Phi,_\s}
\newcommand{\Phikrhonul}{\Phi,_\rho^{(0)}}
\newcommand{\Phikbalph}{\Phi,_\balpha}
\newcommand{\Phikbbet}{\Phi,_\bbeta}
\newcommand{\Phikbmu}{\Phi,_\bmu}
\newcommand{\Phikbnu}{\Phi,_\bnu}

\newcommand{\uijnul}{u_{ij}^{(0)}}
\newcommand{\uijone}{u_{ij}^{(1)}}
\newcommand{\uijtwo}{u_{ij}^{(2)}}
\newcommand{\Uijnul}{U_{ij}^{(0)}}
\newcommand{\Uijone}{U_{ij}^{(1)}}
\newcommand{\Uijtwo}{U_{ij}^{(2)}}
\newcommand{\Uabnul}{U_{\alpha\beta}^{(0)}}
\newcommand{\Uabone}{U_{\alpha\beta}^{(1)}}
\newcommand{\Uabtwo}{U_{\alpha\beta}^{(2)}}
\newcommand{\Unul}{U^{(0)}}
\newcommand{\Urrnul}{U_{rr}^{(0)}}
\newcommand{\Ussnul}{U_{\s\s}^{(0)}}
\newcommand{\Uuunul}{U_{\uc\uc}^{(0)}}
\newcommand{\Ursnul}{U_{r\s}^{(0)}}
\newcommand{\Urunul}{U_{r\uc}^{(0)}}
\newcommand{\Usunul}{U_{\s\uc}^{(0)}}
\newcommand{\Urr}{U_{rr}}
\newcommand{\Uss}{U_{\s\s}}
\newcommand{\Uuu}{U_{\uc\uc}}
\newcommand{\Urs}{U_{r\s}}
\newcommand{\Uru}{U_{r\uc}}
\newcommand{\Usu}{U_{\s\uc}}
\newcommand{\urrnul}{u_{rr}^{(0)}}
\newcommand{\ussnul}{u_{\s\s}^{(0)}}
\newcommand{\uuunul}{u_{\uc\uc}^{(0)}}
\newcommand{\urr}{u_{rr}}
\newcommand{\unn}{u_{nn}}
\newcommand{\uss}{u_{\s\s}}
\newcommand{\uuu}{u_{\uc\uc}}

\newcommand{\linopl}{{\mathcal L}}

\newcommand{\Gammagab}{\Gamma_{\alpha\beta}^{\gamma}}
\newcommand{\Gammagba}{\Gamma_{\beta\alpha}^{\gamma}}

\newcommand{\Gammarrr}{\Gamma_{rr}^{r}}
\newcommand{\Gammarss}{\Gamma_{\s\s}^{r}}
\newcommand{\Gammaruu}{\Gamma_{\uc\uc}^{r}}
\newcommand{\Gammarsr}{\Gamma_{\s r}^{r}}
\newcommand{\Gammarrs}{\Gamma_{r \s}^{r}}
\newcommand{\Gammarur}{\Gamma_{\uc r}^{r}}
\newcommand{\Gammarru}{\Gamma_{r \uc}^{r}}
\newcommand{\Gammarsu}{\Gamma_{\s\uc}^{r}}
\newcommand{\Gammarus}{\Gamma_{\uc\s}^{r}}

\newcommand{\Gammasss}{\Gamma_{\s\s}^{\s}}
\newcommand{\Gammasrr}{\Gamma_{rr}^{\s}}
\newcommand{\Gammasuu}{\Gamma_{\uc\uc}^{\s}}
\newcommand{\Gammassr}{\Gamma_{\s r}^{\s}}
\newcommand{\Gammasrs}{\Gamma_{r \s}^{\s}}
\newcommand{\Gammasur}{\Gamma_{\uc r}^{\s}}
\newcommand{\Gammasru}{\Gamma_{r \uc}^{\s}}
\newcommand{\Gammassu}{\Gamma_{\s\uc}^{\s}}
\newcommand{\Gammasus}{\Gamma_{\uc\s}^{\s}}

\newcommand{\Gammauuu}{\Gamma_{\uc\uc}^{\uc}}
\newcommand{\Gammaurr}{\Gamma_{rr}^{\uc}}
\newcommand{\Gammauss}{\Gamma_{\s\s}^{\uc}}
\newcommand{\Gammausr}{\Gamma_{\s r}^{\uc}}
\newcommand{\Gammaurs}{\Gamma_{r \s}^{\uc}}
\newcommand{\Gammauur}{\Gamma_{\uc r}^{\uc}}
\newcommand{\Gammauru}{\Gamma_{r \uc}^{\uc}}
\newcommand{\Gammausu}{\Gamma_{\s\uc}^{\uc}}
\newcommand{\Gammauus}{\Gamma_{\uc\s}^{\uc}}

\newcommand{\ud}{u}
\newcommand{\potelast}{V_{\rm el}}
\newcommand{\potelastnul}{V_{\rm el}^{(0)}}
\newcommand{\potelastnulbar}{\bar{V}_{\rm el}^{(0)}}
\newcommand{\lamtil}{\tilde{\lambda}}
\newcommand{\mus}{G}
\newcommand{\mustil}{\tilde{G}}
\newcommand{\tnabla}{\tilde{\nabla}}
\newcommand{\hnul}{h_0}

\newcommand{\xvec}{{x}}
\newcommand{\kvec}{{k}}

\newcommand{\jvec}{\mathbf{j}}
\newcommand{\im}{\mathrm{i}}
\newcommand{\driv}{F}
\newcommand{\la}{\Lambda}

\newcommand{\mutil}{{\tilde\mu}}
\newcommand{\muhat}{{\hat\mu}}
\newcommand{\mob}{{\tilde M}}



\renewcommand{\Vec}[1]{{\mathbf #1}}
\section{Introduction}
For a large class of pattern-forming systems, the essential dynamics
to be understood and described is that of an interface between two
phases.  Mathematically speaking, part of the problem to be solved
consists in determining the position of the interface as a function of
time, i.e., is a free or moving-boundary problem.

Phase-field models have been established as powerful tools for the
numerical simulation of this kind of problem. They avoid explicit
front tracking and are versatile enough to deal with topological
changes. Phase-field methods constitute a special case of level-set
approaches \cite{osher88,sethian99}. They differ from the general case
by having a level-set function that satisfies particular bulk
equations of motion rendering unnecessary the computation of an
extension of the interface velocity to the bulk.  This way, they also
avoid interface capturing at each time step, which is normally
requisite in level-set methods.  In comparison with other level-set
methods, a drawback of the phase-field approach is that it does not
yield an exact representation of the interface continuum problem,
reducing to its dynamics only asymptotically.  However, quantitative
numerical control of phase-field representation has made enormous
progress in the last decade \cite{karma96,karma98}, so this
disadvantage is not crucial anymore.

In a phase-field model, information on the interface position is
present implicitly, given either as a level set of a particular value
of the phase field (in two-phase models) or by equality of the
phase-field values for different phases (in multi-phase models), and
can be recovered by computation of the appropriate level set at only
those times when knowledge of the position is desired.

A major field of application are solidification problems, where
diffuse-interface models were developed early on
\cite{langer75,collins85,caginalp86} and have seen renewed interest
ever since computational power increased enough to render their
simulation feasible. The concept was extended to anisotropic interface
properties \cite{mcfadden93}, and first qualitative numerical
calculations of dendritic growth \cite{kobayashi93,kobayashi94} were
followed by theoretical improvement of the asymptotics permitting
quantitative simulations \cite{karma96,karma98}, at least for
intermediate to large undercoolings.  Non-dendritic growth
morphologies were also simulated, even in three dimensions
\cite{abel97}.  Generalizations included the description of the
coexistence of more than two phases \cite{wheeler96,garcke99}.

Additional examples of successful application of the tool phase field
include the modeling of step-flow growth \cite{opl03,voigt04} and of
the elastically induced morphological instability
\cite{mueller99,kassner99,kassner01}, often labeled Grinfeld
\cite{grinfeld86} or Asaro-Tiller-Grinfeld (ATG) instability
\cite{asaro72}.  All of the examples mentioned so far dealt with {\em
  nonconservative} interface dynamics, where a particle reservoir is
provided by either the melt that is in contact with the solid or the
adatom phase on a vicinal surface.

Actually, regarding the ATG instability, which is an instability
with respect to material transport driven by elastic energy, interest
initially focused on transport by surface diffusion, which leads to
{\em conserved} dynamics.  This was the case in the first article by
Asaro and Tiller \cite{asaro72}, but also in the first numerical
simulations by sharp-interface continuum models \cite{yang93},
preceding computations of the instability under transport by
melting-crystallization \cite{kassner94}.

The situation reversed when the phase-field method was employed for
the first time to compute the ATG instability
\cite{mueller99,kassner99}.  Here, all the early works considered a
nonconserved phase-field
\cite{mueller99,kassner99,kassner01,haataja02}. Only recently has
surface diffusion been considered in phase-field models treating
elastically stressed materials \cite{raetz06,yeon06}.
This difference in preferences when modeling either
on the basis of a sharp-interface model or using a phase field may be
due to the fact that writing down a nonconservative and a conservative
model is equally simple in the former case, whereas it is less
straightforward to write down the conservative model within the
phase-field approach than the nonconservative one.

This is not to say that phase-field models with a conservation law for
the phase field have not been considered at all.  Starting from a
Cahn-Hilliard equation with a concentration dependent mobility, Cahn
et al.~\cite{cahn96} obtained an interface equation with the normal
velocity proportional to the Laplacian of the mean curvature.  It then
appears as if all phase-field models with surface diffusion should be
derivable on the basis of similar considerations.  Indeed, comparable
models have been applied in the simulation of electromigration and
voiding in thin metal films \cite{mahadevan99a,bhate00}.  These two
models are slightly different, but the difference is not crucial and
all previous models except the one given in \cite{raetz06} seem to
suffer from the same problem, to be discussed in the following.

As we shall see, it is quite easy to set up a conservative phase-field
model. But it is more difficult to obtain the correct asymptotics
describing surface diffusion as given by the desired sharp-interface
limit. Past models such as the ones presented in
\cite{cahn96,mahadevan99a,bhate00,yeon06} while asymptotically
producing a set of equations \emph{containing} the desired limit
equations, include an \emph{additional restriction}, i.e., they have
one equation too many, a fact that seems to have been overlooked so
far. In \cite{raetz06}, this restriction is not present, but the
authors consider their improvement only a stabilizing element, not
changing the asymptotics, whereas what they have achieved in reality
is superior asymptotic behavior.  Because the flaw of the faulty
models is subtle, it is not a priori clear how adversely the undesired
restriction will affect their behavior.  Therefore, numerical
simulations are necessary to assess their respective virtues and
drawbacks.

The purpose of the present article is to demonstrate the overlooked
restriction, to explore an alternative approach to phase-field
modeling of surface diffusion, to derive additional models not having
the aforementioned flaw, and finally, to compare the different models
numerically.

To render things as simple as possible, we will restrict ourselves to
two dimensions and give analytic derivations only for purely
surface-diffusion-driven motion, i.e., the coupling to a destabilizing
process such as elastic relaxation or electromigration will not be
considered in the asymptotic analysis.  The fully three-dimensional
model including elastic energy and thus describing the ATG instability
has been given in \cite{kassner_condmat06}, an article with lower
pedagogical ambitions than this one. In simulations, we will
consider both surface diffusion and elastic degrees of freedom, i.e.,
the ATG instability, to be able to make comparisons for stable
and unstable situations.

The paper is organized as follows.  In Sec.~\ref{sec:sharp_i_mod}, the
sharp-interface model to be approximated by the phase-field equations
is specified. The nonconservative case will be discussed for reference
purposes. Section \ref{sec:scalar} then presents the standard approach
that previously was supposed to reduce to the correct limit and
pinpoints the oversight in existing asymptotic analyses.  An
alternative approach is presented in Sec.~\ref{sec:tensorial} failing
for complementary reasons.  By appropriate combination of the ideas
from both approaches, two phase-field models will be given in
Sec.~\ref{sec:correct} producing the correct asymptotic behavior.  An
analytic linear stability analysis of these models shows, in
Sec.~\ref{sec:linstab}, that they correctly reproduce the spectra of
the sharp-interface limit.  In Sec.~\ref{sec:numerical}, comparisons
of the different models will be performed via numerical simulation for
a number of pertinent situations. Finally, some conclusions to be
drawn from both analytic and numerical calculations will be discussed.

\section{Sharp-interface model for motion induced by curvature
\label{sec:sharp_i_mod}}

In the simplest case, where surface energy is the only relevant
quantity determining the motion of an interface, the local chemical
potential difference between the solid and the second phase [liquid,
gas (vapour), or vacuum] at the interface may be written
\begin{equation}
\delta \mu  =\frac{1}{\rho_s} \, \gamma \kappa\>,
 \label{eq:delta_mu}
\end{equation}
where $\rho_s$ is the density of the solid phase, $\gamma$ the
(isotropic) surface tension and $\kappa$ the curvature (in 3D, the
mean curvature). A positive curvature corresponds to a locally convex
solid phase, a negative one to a locally concave solid.

Once the chemical potential is given, the stability of the interface
can be assessed.  From \eqref{eq:delta_mu}, we infer immediately that
a planar interface is energetically stable. Any protrusion of the
solid gives a convex bump and increases the energy of the solid over
that of the second phase, leading to a nonequilibrium situation
favoring diminution of the solid phase. Any indentation of the solid
produces a concave bump and decreases the energy of the solid below
that of the second phase, leading to a nonequilibrium situation
favoring growth of the solid phase.

However, in order to determine the evolution of an unstable state,
some dynamical law governing its motion must be stated.  If a particle
reservoir is present and the interface is rough, it is natural to
assume linear nonequilibrium kinetics. The driving force then is the
chemical potential difference itself, and the normal velocity $v_n$ of
the interface will be proportional to it:
\begin{equation}
v_n = - k_v \delta\mu\>,
\label{eq:vn_noncons}
\end{equation}
where $k_v$ is a mobility and the normal points from the solid into
the second phase.

On the other hand, for material transport by surface diffusion, the
driving force is the gradient of the chemical potential along the
surface, producing a surface current $j \propto -\nabla_s
\,\delta \mu$ ($\nabla_s$~is the surface gradient), which leads to a
dynamical law of the form
\begin{equation}
  \label{eq:vn_surfdiff}
  v_n =  M_s \,\Delta_s \delta \mu =  M_s \,
\frac{\partial^2 \delta \mu}{\partial s^2} \>,
\end{equation}
where $\Delta_s$ is the Laplace-Beltrami operator on the surface,
reducing to a double derivative with respect to the arc length for a
one-dimensional interface, and $M_s$ a mobility coefficient
(dimensionally different from the mobility $k_v$), assumed constant
here.

A linear stability analysis of a planar interface is readily
performed, writing
\begin{equation}
\zeta(\xvec,t)=\zeta_0 + \epsilon\zeta_1 e^{\im \kvec \xvec + \omega t}
\label{eq:perturb_plan}
\end{equation}
where $\zeta_0$ is the constant position of the unperturbed interface,
$\xvec$ the cartesian coordinate parallel to it, and $\epsilon$ a
small parameter used to keep track of orders of the perturbation
expansion. The form of the perturbation, containing a wave number
$\kvec$ and a growth rate $\omega$, is dictated by the fact that plane
waves are eigenmodes of the differential operator $\partial_x^2$
appearing in the definition of the curvature \eqref{eq:kappa} and that
$v_n = \dot\zeta/[1+(\partial_x\zeta)^2]^{1/2}$ is proportional to a
first-order time derivative.  Using
\begin{equation}
\kappa = -\frac{\partial_x^2 \zeta}{[1+(\partial_x\zeta)^2]^{3/2}} \>,
\label{eq:kappa}
\end{equation}
one obtains the dispersion relations
\begin{equation}
\omega = - \frac{k_v \gamma}{\rho_s} k^2 \equiv - K k^2
\label{eq:disprel_nc}
\end{equation}
for the nonconservative and
\begin{equation}
\omega = - \frac{M_s \gamma}{\rho_s} k^4 \equiv - M k^4
\label{eq:disprel_co}
\end{equation}
for the conservative cases, respectively.  To arrive at the last
result, note that to linear order (in $\epsilon$) $\partial_s^2$ is
not different from $\partial_x^2$.

For brevity, we have defined new kinetic coefficients $K$ and $M$,
which allows us to avoid carrying along the factor $\gamma/\rho_s$ all
the time.

The two models for motion by curvature considered here are given by
Eqs.~\eqref{eq:delta_mu} and \eqref{eq:vn_noncons} on the one hand and
Eqs.~\eqref{eq:delta_mu} and \eqref{eq:vn_surfdiff} on the other,
describing the nonconservative and conservative cases, respectively.
A phase-field model trying to represent these dynamics should converge
to the appropriate set of these sharp-interface equations in the limit
of asymptotically small interface width. More interesting and more
complex physical problems are obtained, when the normal velocity
depends on additional ingredients beyond curvature-induced driving
forces. This will then lead to a coupling of the phase field to other
fields. We shall discuss such a case later on.


\section{Scalar-mobility phase-field model\label{sec:scalar}}

Before considering the structure of previous phase-field models
attempting to capture surface diffusion dynamics, let us briefly
recall the phase-field model for nonconserved order parameter $\phi$.
This can be written \cite{kassner01}
\begin{equation}
  \label{eq:phidyn_nc}
 \frac{\partial \phi}{\partial t} =
K \biggl( \nabla ^2 \phi -  \frac{2}{W^2} f'(\phi)
\biggr)
\end{equation}
where $f(\phi) = \phi^2 (1-\phi)^2$ is the usual double-well potential
describing two-phase equilibrium. $\phi$ varies between 0,
corresponding to the nonsolid phase, and 1, corresponding to the solid;
$W$ measures the width of the transition region between the two
phases, i.e., it may be interpreted as an interface thickness.
A prime denotes a derivative with respect to the
argument.


The standard approach to a phase-field description of surface
diffusion, as proposed in \cite{yeon06,cahn96,mahadevan99a,bhate00}, is then
to prepend the right hand side of Eq.~\eqref{eq:phidyn_nc} with a
differential operator corresponding to the divergence of a gradient
multiplied by a phase-field dependent mobility, i.e.,
Eq.~\eqref{eq:phidyn_nc} becomes replaced with
\begin{align}
  \frac{\partial \phi}{\partial t} &= \nabla \cdot \jvec   
\nonumber\\
\jvec &= \mob  \nabla \frac{1}{W^2}\delta\mutil(\nabla^2\phi,\phi) \>,
\label{eq:phidyn_c}\\
\delta\mutil(\nabla^2\phi,\phi) &\equiv
-W^2 \nabla^2 \phi+ 2 f'(\phi) \nonumber 
\end{align}
where $\mob $ is a scalar function of either $\phi$ \cite{yeon06,cahn96,bhate00}
or $W\nabla \phi$ \cite{mahadevan99a}, chosen such that the mobility
tends to zero far from the interface: $\mob (\phi,W\nabla\phi)\to 0$ for
$\phi\to 0$ and $\phi\to 1$.  $\delta\mutil$ is a nondimensionalized
chemical potential difference.

We will refer to the model described by Eqs.~\eqref{eq:phidyn_c} as
the \emph{scalar-mobility model} or briefly \emph{SM model} in the following.

At this point, a few remarks are in order.  First, the field $\phi$ is
the density of a conserved quantity by construction, since the right
hand side of \eqref{eq:phidyn_c} is written as a divergence.  This is
true for any (nonsingular) form of the mobility function $\mob $.  Second,
$\delta\mutil$ becomes zero for $\phi\to 0$ and $\phi\to 1$, meaning that
there is no diffusion in the bulk anyway.  One might therefore wonder
whether it is really necessary to choose a mobility that goes to zero
in the bulk.  The conservation law plus the absence of diffusion far
from the interface should suffice to restrict transport to diffusion
along the interface.  In fact, we shall see that essentially the same
asymptotic results are obtained no matter what the form of $\mob $, the
only conditions to be imposed being positivity (for almost all values
of $\phi$ or $\nabla\phi$) and boundedness.  It is just easier to
derive them if it is in addition assumed that $\mob $ vanishes in the
bulk.  On the other hand, it will turn out that if a restriction
imposed by the asymptotics is removed (or not yet satisfied in the
temporal evolution of the system), $\mob $ has to decay sufficiently fast
inside the bulk for the limit to make sense.  This may be relevant for
the behavior of the model before it reaches its asymptotic state.

Finally, the issue at present is not so much whether the dynamics is
conservative but whether it does reduce to the sharp-interface model
of Sec.~\ref{sec:sharp_i_mod} in the limit of an asymptotically
vanishing interface thickness.  To investigate this, we have to
explicitly carry out the asymptotic analysis.

\subsection{Local coordinate system \label{sec:loccoord}}

The basic idea of the analysis is to expand all dynamical quantitities
in terms of the small parameter $W$ describing the interface
thickness, to solve for the phase field and to use the solution to
eliminate its explicit appearance from the equations.  To this end,
the domain of definition of the field is divided into an outer region,
where gradients of fields can be considered to be of order one and
an inner region (close to the interface), where these gradients are of
order $1/W$.  The expansion in powers of $W$ is rather straightforward
in the outer domain, Eq.~\eqref{eq:phidyn_c} can be taken as
a starting point directly.  As to the inner domain (and its matching with the
outer region), it is useful to first transform to coordinates adapted
to its geometry.  Therefore, local coordinates $r$ and $\s$ are
introduced with $r$ orthogonal to the interface (defined as the level
set corresponding to $\phi(x,z,t)=1/2$) and $\s$ tangential to it.
$r$ is the signed distance from the interface and will be rescaled by
a stretching transformation $r=W\rho$ to make explicit the $W$
dependence for the expansion, $\s$ is the arc length of the interface
curve.  Inner and outer solutions must satisfy certain matching
conditions due to the requirement that they agree in the combined
limit $W\to 0$, $\rho\to\pm\infty$, $r\to 0$.  These conditions are
given in App.~\ref{sec:matching}.

To obtain a set of basis vectors for our local coordinate system, we
first write
\begin{equation}
  \label{eq:posvec_loc}
 \rvec = \Rvec(\s)+r\, \norm(\s)\>,
\end{equation}
where $\rvec$ is the position vector of a point near the interface,
$\Rvec$ the position of the interface itself, and $\norm$ the normal
vector on it (oriented the same way as in the sharp-interface model,
i.e., pointing out of the solid).

Given the coordinates, it is a trivial matter to write down a
coordinate basis
\begin{equation}
 \label{eq:covar_bas}
\begin{aligned}
\Er  &\equiv \frac{\partial\rvec}{\partial r} = \norm(\s)  \>,
  \\
\Es  &\equiv \frac{\partial\rvec}{\partial \s} =
\frac{\partial\Rvec}{\partial \s}
+ r \frac{\partial\norm}{\partial \s} =  \left(1+r \kappa\right)\tang
 \>,
\end{aligned}
\end{equation}
which is orthogonal. (This is no longer automatically true in 3D
\cite{kassner_condmat06}.) $\tang = \partial\Rvec/\partial \s$ is the
unit tangent vector to the interface, and $\partial\norm/\partial \s =
\kappa \tang$ is one of the Frenet formulas \cite{spivak79},
specialized to two dimensions.

From \eqref{eq:covar_bas}, we first obtain the metric coefficients
$g_{\alpha\beta}={\mathcal E}_\alpha {\mathcal E}_\beta$, where
$\alpha,\beta \in \left\{r,\s\right\}$. The metric tensor
reads
\begin{align}
\label{eq:metric}
& \left(g_{\alpha\beta}\right) =  {\mathbf g} =
 \left(\begin{array}{cc}
1 &              0   \\
0 & (1+r \kappa)^2 \\
\end{array} \right) ,
\end{align}
its determinant is
\begin{equation}
g \equiv \det{\mathbf g} = \left(1+ r \kappa\right)^2 \>,
\end{equation}
hence $\sqrt{g} = 1+ r \kappa$ (using the locality of $r$ to ascertain
$r\kappa<1$), and the contravariant components of the metric tensor
are obtained as
\begin{align}
\label{eq:inv_metric}
& \left(g^{\alpha\beta}\right) =    {\mathbf g}^{-1}
 =  \left(\begin{array}{cc}
1 &              0   \\
0 & (1+r \kappa)^{-2}
\end{array} \right) .
\end{align}

From now on, we use the Einstein summation convention for pairs of
covariant and contravariant indices.  The vectors of the reciprocal
basis are obtained from $\Ehalph = g^{\alpha\beta} \Ebet$:
\begin{equation} \label{eq:contravar_bas}
\begin{aligned}
\Ehr  &= \nabla r = \norm(\s)  \>,   \\
\Ehs  &= \nabla \s = \frac{1}{\sqrt{g}} \tang \>,
\end{aligned}
\end{equation}
The gradient and divergence read
\begin{align}
\nabla &= \Ehalph \partial_\alpha \>, \label{eq:nabla}\\
\nabla\cdot \Avec &= \frac{1}{\sqrt{g}} \partial_\alpha
\left(\sqrt{g} g^{\alpha\beta} A_\beta\right) \>.
\label{eq:divergence}
\end{align}
Note that on the interface, the covariant component $A_r$ is equal to
the normal component $A_n$, but that $A_s$ is related to the
tangential component $A_t$ by $A_s= \sqrt{g} A_t$, because $\Ehs$ is
not normalized to one.

In the following, we will denote inner quantities by the uppercase
letter corresponding to the lowercase letter denoting the outer quantity,
whenever this is meaningful.

Since the interface will move in general and the coordinates $r$ and
$\s$ are defined with respect to the interface,
there is also a transformation rule for the time derivative:
\begin{equation}
\partial_t f(x,z,t) = \partial_t F(r,\s,t) -
\wel \nabla F(r,\s,t) \>,
\label{eq:timederiv}
\end{equation}
where $\wel(\s,t)$ is the
interface velocity. Equation \eqref{eq:timederiv} exhibits that the
time derivative in the comoving frame is a material derivative. In
order to formulate the matching conditions concisely, we will
occasionally also write the outer fields as functions of the variables
$r$ and $\s$ (without changing their naming letter, thus in this
case adhering to the physicists' convention of using a letter for a
physical quantity rather than a mathematical function).

\subsection{Inner equations}

To render the scales of the different terms more visible, we rewrite
Eqs.~\eqref{eq:nabla} and \eqref{eq:divergence} in the
form
\begin{align}
\nabla & = \frac{1}{W} \,\norm\, \partial_\rho
+ \frac{1}{\sqrt{g}}\, \tang\,  \partial_\s \>, \label{eq:nablaeps}\\
\nabla\cdot \Avec & =
  \frac{1}{\sqrt{g}} \left(\frac{1}{W}\,\partial_\rho \sqrt{g} A_r +
 \partial_\s
\frac{1}{\sqrt{g}}\, A_\s \right) \>,
\label{eq:divergenceeps}\\
\sqrt{g} & = 1+ W\rho\kappa  \>. \label{eq:sqrtg}
\end{align}

Assuming, without loss of generality, that the tangential velocity of
the interface vanishes, Eq.~\eqref{eq:phidyn_c} takes the following
form
\begin{align}
  \partial_t \Phi -\frac{1}{W} v_n \partial_\rho \Phi = &
\nabla \cdot \mob  \nabla \frac{1}{W^2}\,
\delta\mutil(\nabla^2\Phi,\Phi)\>,\nonumber\\
 \delta\mutil(\nabla^2\Phi,\Phi) = &
-\frac{1}{\sqrt{g}}\,\partial_\rho \sqrt{g}\,\partial_\rho
 \Phi  \label{eq:phidyn_c_inn}\\
 & - W^2\frac{1}{\sqrt{g}}\,\partial_\s \frac{1}{\sqrt{g}}\,
 \partial_\s \Phi
 + 2 f'(\Phi)
\>,\nonumber
\end{align}

with 
\begin{align}
  \label{eq:diffop1_inn}
\nabla \cdot \mob  \nabla &= \frac{1}{W^2} \frac{1}{\sqrt{g}}\partial_\rho
 \sqrt{g} \mob  \partial_\rho   + \frac{1}{\sqrt{g}}\,\partial_\s \frac{1}{\sqrt{g}}\,
\mob  \partial_\s \>.
\end{align}
Hence, the leading term of the inner equation \eqref{eq:phidyn_c_inn}
with the differential operator given by \eqref{eq:diffop1_inn} is of
order $W^{-4}$.

\subsection{Expansions, matched asymptotic analysis \label{sec:expansions}}

To solve the outer and inner equations successively, we expand the phase
field in both the outer and inner domains in powers of $W$
\begin{eqnarray}
\phi(x,z,t) &=& \phinul(x,z,t) + W\, \phione(x,z,t)
\nonumber\\
 && \mbox{} + W^2\phitwo(x,z,t)...
           \label{phioutexp}
\end{eqnarray}
and
\begin{eqnarray}
\Phi(r,\s,t) &=& \Phinul(r,\s,t) + W\,
\Phione(r,\s,t)
\nonumber\\
 && \mbox{} + W^2\Phitwo(r,\s,t)... \>.
           \label{phiinnexp}
\end{eqnarray}
We now proceed solving the outer and inner equations order by order.

\subsubsection{Leading order \label{sec:expleading}}

The leading-order outer equation for $\phi$ is
\begin{equation}
\label{eq:scalout_0}
 \nabla \cdot \mob  \nabla f'(\phinul) = 0\>,
\end{equation}
which is to be supplemented with the boundary conditions $\phinul=1$
and $\phinul=0$ at infinity in the regions where the system is solid
and non-solid, respectively.  If we regard \eqref{eq:scalout_0} as a
partial differential equation for the function $f'(\phinul)$ (rather
than for $\phinul$ itself), this boundary condition translates into
$f'(\phinul)\to 0$ as $\abs{\rvec}\to \infty$, which may be seen
immediately from the explicit form of $f'(\phi)$, given in
App.~\ref{sec:collection}.  The new boundary condition is valid
everywhere at infinity except possibly in a region with size of order
$W$.  For general $\mob (\phi,W\nabla\phi)$, the partial
differential equation (\ref{eq:scalout_0}) is of course nonlinear.
Nevertheless, it can be shown to have the unique solution $f'(\phinul)
= 0$, if $\mob $ is positive everywhere, except possibly on a set of
measure zero.

To see this, multiply Eq.~\eqref{eq:scalout_0} by $f'(\phinul)$,
integrate over all of space and use Gauss's theorem:
\begin{align}
\label{eq:uniqueness}
 0 &= \int dV f'(\phinul) \nabla \cdot \mob  \nabla f'(\phinul)
 \nonumber\\
 &= -\int dV  \mob  \left[\nabla f'(\phinul)\right]^2\> + O(W)\>,
\end{align}
where the $O(W)$ stands for the surface integral at infinity. (We will
often use the nomenclature for three-dimensional systems, not
distinguishing between surface integrals and boundary contour
integrals; also we employ $dV$ for the volume element of both two- and
three-dimensional space.)  If $\mob $ is positive almost everywhere, we
immediately get $f'(\phinul)=\hbox{const.}$, and the boundary
conditions require the constant to be zero.  This conclusion remains
of course unchanged, if $\mob $ becomes zero only when $\phinul$ is zero
or one -- a standard choice \cite{bhate00} is $\mob (\phi) \propto \phi^2
(1-\phi)^2$.

Hence, the unique solution to the leading-order outer problem is, if
we now consider it an equation for $\phinul$ again, $\phinul=1$ in
$\Omega^{-}$ and $\phinul=0$ in $\Omega^{+}$, where $\Omega^{\mp}$ are
those regions of space, separated by the interface(s), in which
$\lim_{\abs{\rvec}\to\infty} \phinul = 1$ and $0$, respectively.  The
solution $\phinul=\frac12$, still possible for the equation
interpreted as an equation for $f'(\phinul)$, is excluded by the
boundary conditions for $\phinul$.  (This argument presupposes that we
have no domains that are not connected with infinity.  For the
interior of a closed interface, the solution $\phinul=1/2$ would have
to be excluded by a stability argument or by making reference to
initial conditions.)

It is then seen by inspection that the outer equation is indeed solved
to {\em all\/} orders by the solution under discussion.  With the
usual construction of coupling terms to, say, mechanical or electrical
degrees of freedom \cite{kassner99,mahadevan99a}, this remains true
for the full phase-field model including the coupling, as these terms
are typically made to vanish in the bulk. So our statement, which has
some importance as we shall see, is valid beyond the oversimplified
``free'' model considered here for the purpose of demonstration. Any
deviation of the outer solution from $\phinul$ must be
transcendentally small.

Therefore, we have $\phione\equiv0$, $\phitwo\equiv0$, which provides
us with partial boundary conditions for the inner solutions $\Phione$,
$\Phitwo$, and so on (see App.~\ref{sec:matching}).  Moreover, only
the inner problem needs to be considered beyond the leading order.

Because $g=1+O(W)$, the leading-order inner problem becomes [see
Eqs.~\eqref{eq:phidyn_c_inn} and \eqref{eq:diffop1_inn}]
\begin{equation}
\label{eq:phydin_c_innlead}
\partial_\rho \mob (\Phinul) \partial_\rho \left[\partial_{\rho\rho} \Phinul -2
f'(\Phinul)\right]=0 \>,
\end{equation}
where $\partial_{\rho\rho}=\partial_\rho^2$.
This can be integrated once to yield
\begin{equation}
\label{eq:phydin_c_innlead_int}
\partial_\rho \left[\partial_{\rho\rho} \Phinul -2
f'(\Phinul)\right] = \frac{c_1(\s)}{\mob (\Phinul)} \>,
\end{equation}
where $c_1(\s)$ is a function of integration.  It is here that we
have to follow different lines of arguments, depending on whether $\mob $
approaches zero for $\rho\to\pm\infty$, which is the case for the
mobilities assumed in \cite{yeon06,cahn96,bhate00}, or whether it is just a
bounded (and possibly constant) function of $\phi$.  In the first
case, we may immediately conclude $c_1=0$, because the right hand side
of (\ref{eq:phydin_c_innlead_int}) must not diverge.  In the second
case, we obtain the same result by integrating
\eqref{eq:phydin_c_innlead_int} first and invoking the boundary
conditions:
\begin{equation}
\label{eq:phydin_c_innlead_int2}
\partial_{\rho\rho} \Phinul -2
f'(\Phinul) = c_1(\s)\int_0^\rho \frac{1}{\mob } \,d\rho + c_2(\s)\>.
\end{equation}
Since $\mob $ is bounded from above and positive, the integral will be
larger in magnitude than $\int_0^\rho 1/(\sup_\rho \mob ) d\rho
=\rho/\sup_\rho \mob $, so the two factors multiplying $c_1$ and $c_2$
are linearly independent.  The left hand side approaches zero for
$\rho\to\pm\infty$ [the argument will be made more rigorous below in
the discussion of $\Phione$], so both $c_1$ and $c_2$ must be equal to
zero.  To argue that $c_2$ is zero in the case where $\mob \to0$ for
$\rho\to\pm\infty$, we can proceed the same way, except that we have
already gotten rid of the term containing $c_1$, so the right hand
side of \eqref{eq:phydin_c_innlead_int2} is $c_2$ only.

To summarize, the leading-order inner equation results in
\begin{equation}
\label{eq:phidyn_c_innlead_res}
\partial_{\rho\rho} \Phinul -2 f'(\Phinul) = 0\>,
\end{equation}
and this provides us with the solution $\Phinul= \frac12
\,(1-\tanh\rho)$ as is shown in App.~\ref{sec:collection}.

\subsubsection{Next-to-leading order}

The next-to-leading order in Eq.~\eqref{eq:phidyn_c_inn} is the order
$W^{-3}$.  Since the differential operator in front of the
chemical potential is of order $W^{-2}$ and the chemical
potential multiplied by another factor $W^{-2}$, we must expand
$\delta\mutil$ up to order $W$.  Equation
\eqref{eq:phidyn_c_innlead_res} already tells us that
$\delta\mutil^{(0)}=0$, so we obtain
\begin{equation}
\label{eq:phydin_c_innnext}
\partial_\rho \mob (\Phinul) \partial_\rho \delta\mutil^{(1)}=0 \>,
\end{equation}
from which we get
\begin{equation}
\label{eq:phydin_c_innnext1}
 \partial_\rho \delta\mutil^{(1)}= \frac{d_1(\s)}{\mob (\Phinul)} \>.
\end{equation}
As before, we can immediately conclude from this that $d_1=0$, if we
assume $\mob (\Phinul)\to 0$ for $\Phinul\to 0,1$. For arbitrary but bounded
$\mob $, we invoke the matching conditions (see App.~\ref{sec:matching})
\begin{equation}
\lim_{\rho\to\pm\infty} \partial_\rho \delta\mutil^{(1)} = \partial_r
 \delta\mutil^{(0)}_{\rm out}\vert_{\pm 0} = 0
\end{equation}
to obtain the same result (where for once we have denoted an outer
quantity by a subscript ''out'').

Integrating once more with respect to $\rho$ and writing out
$\delta\mutil^{(1)}$, we have
\begin{eqnarray}
\label{eq:deltamu1res}
  \delta\mutil^{(1)} &=& -\partial_{\rho\rho} \Phione - \kappa
 \partial_\rho \Phinul
+ 2 f''(\Phinul)\,\Phione \nonumber\\
 &=& d_2(\s) \>.
\end{eqnarray}
Up to this point, there is agreement between this and preceding
asymptotic analyses \cite{mahadevan99a}, if not in all details
of the procedure, so at least in the results.

Let us now try to determine the function of integration $d_2(\s)$.
A priori, there is no reason to use a procedure different from what we
have done before.  We know the limiting behavior for $\rho\to\pm\infty$
for two of the four terms on the right hand side~of \eqref{eq:deltamu1res}:
$\lim_{\rho\to\pm\infty} \partial_\rho \Phinul = 0$ [which follows
from either the matching conditions or by inspection of the solution
\eqref{eq:solution_Phi0}] and $\lim_{\rho\to\pm\infty} d_2(\s) =
d_2(\s)$ (because $d_2$ is independent of $\rho$).  Moreover, from
the matching conditions, we obtain the limit for $\Phione$
\begin{eqnarray}
\label{asympt_phione}
\Phione \sim \rho\phi'^{(0)}(\pm0)+\phione(\pm0)
&=& \phione(\pm0) = 0 \nonumber\\
&&(\rho\to\pm\infty) \>.
\end{eqnarray}
The second equality follows from the fact that $\phinul=0$ or
$\phinul=1$, hence its derivative with respect to $r$ vanishes on both
sides of the interface; the third equality is a consequence of the
fact that $\phinul$ solves the outer equation to all orders and hence
$\phione\equiv 0$. In addition, it can be shown \cite{kassner_condmat06}
that if the interaction with mechanical degrees of freedom is
included, this will only lead to terms that also vanish in the limit
$\rho\to\pm\infty$.

With three of the four terms in \eqref{eq:deltamu1res} having a
definite limit, we may conclude that the fourth must have a limit
as well and obtain
\begin{equation}
\label{eq:limitd2phi1}
\lim_{\rho\to\pm\infty} -\partial_{\rho\rho} \Phione = d_2(\s)\>.
\end{equation}
But if this limit exists, it cannot be different from zero:
transforming to $\xi=1/\rho$, we see that $\partial_{\rho\rho} \Phione
= \big(\xi^2 \partial_\xi \big)^2 \Phione$, which implies the
asymptotic behavior $\Phione\sim -d_2/2\xi^2\>\> (\xi\to 0)$ and
hence the divergence of $\Phione$ as $ -d_2 \rho^2/2$, if $d_2\ne 0$.
(The same kind of argument can be used to show that the left hand side
of Eq.~\eqref{eq:phydin_c_innlead_int2} goes to zero, even though the
matching conditions do not provide a direct expression for
$\lim_{\rho\to\pm\infty}\partial_{\rho\rho} \Phinul$.)

The upshot of these detailed considerations is that
\begin{equation}
  \label{eq:mu1eq0}
d_2(\s) = \delta\mutil^{(1)} = 0.
\end{equation}
Previous treatments of the problem did not enter into these
considerations.  Instead, one of the  following two equivalent approaches
was chosen.  Either, Eq.~\eqref{eq:deltamu1res} was interpreted as a
linear inhomogeneous differential equation for $\Phione$ and
Fredholm's alternative invoked.  Since the appearing linear
operator
\begin{equation}
  \label{eq:linop}
 \linopl = \partial_{\rho\rho}-2 f''(\Phinul)
\end{equation}
is hermitean, we know (from taking the derivative of
Eq.~(\ref{eq:phidyn_c_innlead_res}) w.r.t.~$\rho$) that
$\partial_\rho\Phinul$ is a solution to the adjoint homogeneous
equation. The inhomogeneity of the differential equation must be
orthogonal to this solution.  Or else, Fredholm's alternative was not
mentioned, the equation was simply multiplied by
$\partial_\rho\Phinul$, integrated, and it was shown via integration
by parts that the terms containing $\Phione$ disappear. Of course,
this is the same thing.

We then obtain from \eqref{eq:deltamu1res}
\begin{align}
\label{eq:fredholm_1}
& -\int_{-\infty}^{\infty} \kappa \left(\partial_\rho
\Phinul\right)^2 d\rho
= \int_{-\infty}^{\infty} \partial_\rho \Phinul d_2(\s) d\rho = -
d_2(\s) \>,
\end{align}
from which we get, using \eqref{eq:intdphirho2},
\begin{equation}
  \label{eq:fredholm_2}
   d_2(\s) =  \frac13 \kappa \>.
\end{equation}

Both Eqs.~\eqref{eq:mu1eq0} and \eqref{eq:fredholm_2} were derived by
valid methods, therefore they should both hold true.  Nevertheless, as
we shall see shortly, Eq.~\eqref{eq:mu1eq0} is a quite undesirable
result.  This may be the deeper reason why it was so far overlooked
and only the analog of Eq.~\eqref{eq:fredholm_2} derived.  When
Eq.~\eqref{eq:mu1eq0} is inserted in \eqref{eq:fredholm_2}, it leads
to zero curvature at lowest order. In a phase-field model for the ATG
instability \cite{kassner_condmat06}, the same kind of reasoning
imposes a relationship between the elastic state of the material and
the curvature.
In models, where the interaction term is quadratic in $W$
\cite{mahadevan99a}, it again imposes the restriction $\kappa=O(W)$.
To summarize, in all cases we obtain a restriction on the curvature at
lowest order, which means that the phase-field model will not be
asymptotic as long as the deviation from this imposed value is not small.

\subsubsection{Higher orders\label{sec:highorder}}
To see that the model would indeed work if we did not have the
restriction \eqref{eq:mu1eq0}, let us consider the equations at the
next two orders, ignoring for the time being the result $d_2=0$.
Since both $\delta\mutil^{(0)} (=0)$ and $\delta\mutil^{(1)}$ are
independent of $\rho$, the first term of the operator
\eqref{eq:diffop1_inn} does not produce any contribution from these
terms in (\ref{eq:phidyn_c_inn}), and the order $W^{-2}$
equation reads
\begin{equation}
  \label{eq:phydin_c_inn2}
  \partial_\rho \mob  \partial_\rho \delta\mutil^{(2)}
   +\partial_\s   \mob  \partial_\s\,\delta\mutil^{(0)}
= 0 \>,
\end{equation}
where we can immediately drop the second term, because of
$\delta\mutil^{(0)} =0$.  After two integrations this becomes
\begin{equation}
  \label{eq:inn2_int}
 \delta\mutil^{(2)} =
  e_1(\s)\int_0^\rho \frac{1}{\mob } d\rho + e_2(\s)\> \>.
\end{equation}
If $\mob \to 0$ for $\rho\to\pm\infty$, we immediately find $ e_1(\s)=0$.
 In the general case, we use the matching conditions [see \eqref{eq:asymrels3}]
\begin{eqnarray}
  \label{eq:matchmu2}
  \delta\mutil^{(2)} &\sim& \frac12 \rho^2 \,\partial_{rr}
\delta\mutil^{(0)}_{\rm out}\vert_{r=\pm0}
+ \rho \,\partial_{r} \delta\mutil^{(1)}_{\rm out}\vert_{r=\pm0} \nonumber\\
&&\mbox{} +  \delta\mutil^{(2)}_{\rm out}\vert_{r=\pm0} \>.
\end{eqnarray}
From Eq.~\eqref{eq:phidyn_c}, we gather that an expansion of
$\delta\mutil_{\rm out}$ in powers of $W$ will contain three types
of terms, the first of which have the form $\nabla^2 \phi^{(k)}$
($k=0,1,\dots$), while the second contain factors $\phi^{(k)}$
($k=1,2, \dots$), coming from an expansion of $f'(\phi)$
about $\phinul$, and the third include $f'(\phinul)$ alone.
All of these terms vanish, because $\phi^{(k)}=0$ for $k>0$ and
because $f'(\phinul) = 0$.  This is simply a consequence
of the fact that the outer equation is solved exactly by $\phinul =0$
and $\phinul = 1$.  The ``chemical potential'' appearing in the
phase-field equations needs to be related to the true, i.e.,
sharp-interface chemical potential only at the interface.  In the
outer domain, it is zero.  We can then conclude from
\eqref{eq:matchmu2} that $e_1(\s)=0$ (of course $e_2(\s)=0$, too, but
we shall not make use of that result).

Given that $\delta\mutil^{(2)}$ is independent of $\rho$, the inner
equation at order $W^{-1}$ takes the form
\begin{equation}
  \label{eq:phydin_c_inn3}
  -v_n \partial_\rho \Phinul =
\partial_\rho \mob  \partial_\rho \delta\mutil^{(3)}
   +\partial_\s   \mob  \partial_\s\,\delta\mutil^{(1)} \>.
\end{equation}
 After integration over $\rho$
($v_n$ does not depend on $\rho$) we find
\begin{eqnarray}
   \label{eq:veloc_scalmodprel}
   v_n &=& \partial_{\s\s}\delta\mutil^{(1)}\!
\int_{-\infty}^{\infty}\!\mob (\Phinul)\,d\rho +  \mob (\Phinul)\,
\partial_\rho\delta\mutil^{(3)}\Big\vert_{-\infty}^{\infty}\,.
\end{eqnarray}
Here we can drop the second term on the right hand side, if
$\lim_{\rho\to\pm\infty}\mob (\Phinul)=0$.  Formally setting
$\int_{-\infty}^{\infty} \mob (\Phinul)\, d\rho = 3 M$ and using
\eqref{eq:fredholm_2}, we arrive at
\begin{equation}
 \label{eq:veloc_scalmod}
 v_n =  M \partial_{\s\s}\kappa \>.
\end{equation}
Hence, \eqref{eq:veloc_scalmod} reproduces the sharp-interface limit
\eqref{eq:vn_surfdiff}, with the relationship between $M_s$ and $M$
defined in \eqref{eq:disprel_co}.

Finally, $M$ would be infinite for positive functions
$\mob(\Phinul)$ that do not reduce to zero for $\rho\to\pm\infty$;
therefore, in the end we would indeed have to require $\mob(\Phinul)$ to
decay far from the interface, if $\delta\mutil^{(1)}$ were different from
zero.  In reality, we do not just have \eqref{eq:veloc_scalmod}, the
equation we want, but in addition Eq.~\eqref{eq:mu1eq0}, requiring
$\delta\mutil^{(1)}=0$ and, consequently
\begin{equation}
 \label{eq:veloceq0}
 v_n =  0 \>.
\end{equation}

At first sight, Eqs.~\eqref{eq:veloc_scalmod} and
\eqref{eq:veloceq0} may look like contradicting each other, as we can
prepare an initial state with arbitrary curvature of the interface, and
hence the velocity should be different from zero according to
\eqref{eq:veloc_scalmod} but equal to zero according to
\eqref{eq:veloceq0}. However, in preparing an arbitrary initial state,
we have no certainty that the system will already follow its
(lowest-order) asymptotic dynamics. A similar phenomenon happens in
\emph{all} phase-field models when a simulation is started with an initial
interface perturbed by white noise.  Since the asymptotics of the
phase-field equations require curvatures to be smaller than $1/W$, the
initial stage of the dynamics where larger curvatures are present,
will not be governed by these asymptotics.  But in that case the
asymptotic behavior is sufficiently robust to keep the initial stage
short.

Since \eqref{eq:veloceq0} is a lowest-order result for the interface
velocity, we can conceive of two different scenarios.  Either the
next-order result for $v_n$ is nonzero.  Then the model might
asymptotically reproduce the desired sharp-interface limit at the next
order, but its utility would be restricted as it would be quantitative
only for $\kappa = O(W)$ (i.e., $\kappa W \ll W/L$, where $L$ is a
typical system length scale such as, for example, the equilibrium
diameter of a crystal).  Its validity would be restricted to
near-equilibrium situations.  Or else $v_n$ is zero at all orders of
the asymptotic expansion, hence transcendentally small in the
asymptotic limit.  Again, this could describe a near-equilibrium
situation at best. Moreover, such a model would violate the spirit of
phase-field models in general, in which we seek to have an analytic
statement about the sharp-interface limit with as short an expansion
as possible.  The necessity to perform asymptotics beyond all orders
should not arise in problems where we have such a high degree of
freedom in constructing the model equations.

As to the general behavior of the SM model, it is quite tempting to
speculate that when conditions are such that \eqref{eq:veloceq0} does
not hold yet, the phase-field model discussed here will satisfy all
the other less restrictive conditions already, including
\eqref{eq:veloc_scalmod}.
Then the model would be applicable during the period where the
influence of condition \eqref{eq:mu1eq0} leading to
\eqref{eq:veloceq0} is still small.  However, it should be clear that
without a theoretical estimate of the error in this not fully
asymptotic state, the model can hardly be considered quantitative.
Condition \eqref{eq:mu1eq0} should be expected to have a stabilizing
influence on decaying modes of the interface, accelerating their
relaxation towards equilibrium.  Its effect on growing, i.e.~unstable
modes is difficult to assess.

It is instructive to note why the nonconservative model obtained when
\eqref{eq:phidyn_c} is replaced with \eqref{eq:phidyn_nc} does {\em not\/}
suffer from a similar difficulty.  In that model, the velocity is already
determined at the next-to leading order.  Instead of
\eqref{eq:deltamu1res}, we get
\begin{eqnarray}
\label{eq:veloc_nc}
-v_n \partial_\rho \Phinul&=&
K \bigg\{\partial_{\rho\rho} \Phione
 + \kappa \partial_\rho \Phinul
 - 2 f''(\Phinul)\,\Phione \bigg\}
 \>. \nonumber\\
\end{eqnarray}
Again we may conclude that all the terms on the right hand side go to
zero as $\rho$ is sent to $\pm\infty$.  However, this does not lead to
any constraints, since the left hand side is $\rho$ dependent now and
goes to zero as well, satisfying the limit automatically, whereas in
the surface-diffusion case, it was a function of $\s$  only
($d_2$) that could be concluded to be equal to zero.  So consideration
of the limit does not produce anything new here, and the only
procedure available to extract information on $v_n$ is to use
Fredholm's alternative which gives the correct sharp-interface limit.

In the case of the nonconservative model, the introduced chemical
potential functional is zero in the bulk just as in the conservative
case, but there are no restrictions on its variation near the
interface, where it acquires a form tending to a $\delta$ function in
the sharp-interface limit.  In the conservative model, this is
excluded by restrictions on the derivative of the chemical potential
with respect to $\rho$, meaning that the latter must be smooth across
the interface.  Since it is zero off the interface, it is zero on it
as well.  Due to this reason, the phase-field model strictly speaking
applies only to the equilibrium limit.  Far-from equilibrium dynamics
is not likely to be captured faithfully.

Out of the phase-field models for surface diffusion considered in the
literature, the only one that is (apart from our own work
\cite{kassner_condmat06}) not subject to the criticism offered here
seems to be the one given by R\"atz, Ribalta, and
Voigt~\cite{raetz06}. Let us briefly discuss the asymptotics of this
model that we will henceforth denote as the \emph{RRV model}.  In
their simplest form, i.e., for isotropic surface tension and vanishing
kinetic coefficient, the model equations read
\begin{equation}
  \label{eq:phidyn_c_voigt}
\begin{aligned}
  \frac{\partial \phi}{\partial t} &= \nabla \cdot \jvec \\
\jvec &= M B(\phi) \nabla \frac{1}{W^2}\delta\muhat(\nabla^2\phi,\phi) \>,
  \\
\delta\muhat &= \frac{1}{g(\phi)}\left(
-W^2 \nabla^2 \phi+  2 f'(\phi)\right)\,,
\end{aligned}
\end{equation}
with mobility function $B(\phi)=12 \phi^2 (1-\phi)^2$, double-well
potential $f(\phi)= \phi^2 (1-\phi)^2$, and the so-called stabilizing
function $g(\phi)=10 \phi^2 (1-\phi)^2$. Here, we have rescaled the
equations from \cite{raetz06} so as to obtain the same zeroth-order
interface profile as in the SM model (with the original equations,
the interface would have one third of the width ouf our profile).
The leading-order inner problem becomes \eqref{eq:phidyn_c_innlead_res} again.
 At next-to
leading order, we obtain  $\delta\muhat^{(1)}=d_2(s)$ (as before), but
now the chemical potential function is defined differently -- it has a
prefactor that diverges in the bulk
\begin{eqnarray}
\label{eq:deltamu1alt}
  \delta\muhat^{(1)} &=&
 \frac{1}{g(\Phinul)}\left( -\partial_{\rho\rho} \Phione - \kappa
 \partial_\rho \Phinul+ 2 f''(\Phinul)\,\Phione\right) \>.\nonumber\\
\end{eqnarray}
(The first-order term due to variation of the denominator vanishes, as
it contains the differential expression from the left hand side of
\eqref{eq:phidyn_c_innlead_res} as a factor.) The numerator of the
right hand side of \eqref{eq:deltamu1alt} goes to zero as
$\rho\to\pm\infty$ but so does the denominator $g(\Phinul)$, which
renders $\delta\muhat^{(1)}$ indefinite, thus introducing the degree of
freedom necessary for a nonzero value $d_2(\s)$.
Multiplying the equation by $g(\Phinul)\partial_\rho\,\Phinul$ and integrating
with respect to $\rho$ from $-\infty$ to $\infty$, we arrive at
\begin{equation}
\label{eq:fredholm2}
d_2 \int_{-\infty}^{\infty} g(\Phinul)\, \partial_\rho\Phinul \, d\rho
= - \kappa \int_{-\infty}^{\infty} \left(\partial_\rho\Phinul\right)^2 \,
d\rho \>,
\end{equation}
where use has been made of the fact that $\Phinul$ is a left null
eigenvector of the linear operator [inside the parentheses in
\eqref{eq:deltamu1alt}] acting on $\Phione$, to get rid of the
$\Phione$ terms.  The integrals in \eqref{eq:fredholm2} are evaluated
in App.~\ref{sec:collection}, they are equal to -1/3 and 1/3,
respectively.  Hence, $d_2 = \kappa$.

The steps for the following two orders of $W$ follow precisely the
scheme leading from \eqref{eq:phydin_c_inn2} to
\eqref{eq:veloc_scalmod}, which then yields $v_n=M^\ast \partial_{\s\s}
\kappa$, where $M^\ast=\int_{-\infty}^{\infty} M
B(\Phinul)\,d\rho = M$ (for the integral see
App.~\ref{sec:collection}), hence we obtain the desired
sharp-interface limit \eqref{eq:vn_surfdiff}.

Note that with this model, it is essential that the mobility function
goes to zero off the interface.  For the chemical potential
$\delta\muhat$ varies in the bulk (it behaves as $d_2(s)$ near the
interface), hence diffusion there must be suppressed by a vanishing
mobility.

\section{Tensorial mobility\label{sec:tensorial}}
While the RRV model avoids the mistake of imposing \eqref{eq:mu1eq0},
it does so by a purely mathematical device, the introduction of the
stabilizing function $g(\phi)$.  It is then natural to ask whether an
accurate model may not be derived on the basis of mainly physical
considerations.

That the phase-field model given by Eq.~\eqref{eq:phidyn_c} does not
quite yield the correct asymptotics may be traced back to the fact
that the differential operator $\nabla\cdot \mob \nabla$, prepended to
the chemical potential, does not reduce to the surface Laplacian
$\Delta_s$ in the asymptotic limit.  In fact, the second term on the
right hand side of Eq.~\eqref{eq:diffop1_inn} {\em is}, up to a
factor, the Laplace-Beltrami operator on the surface (for $\rho=0$),
but the first term, containing derivatives with respect to $\rho$ is
orders of magnitude larger, being preceded by a factor of $1/W^2$.  As
a consequence, the asymptotics must be secured by the full solution of
the equation rather than by both the operator and the chemical
potential converging to the desired sharp-interface limits.

Realizing this property of the model, it seems natural to modify the
differential operator via introduction of an essentially tensorial
mobility.  Let us denote by
\begin{equation}
  \label{eq:normhat}
  \normhat = -\frac{\nabla\phi}{\abs{\nabla\phi}}
\end{equation}
the normal on the surface $\phi={\rm const.}$ (for $\phi=1/2$, we
have $\normhat=\norm$), then we expect the operator $\nabla\cdot P \nabla$ with
\begin{equation}
  \label{eq:diffop_proj}
  P= 1-\normhat:\normhat
\end{equation}
(cartesian components: $P_{ij}=\delta_{ij}-\hat{n}_i\hat{n}_j$) to reduce to
the surface Laplacian asymptotically. A colon is used to de\-signate a
dyadic product, so $P$ is a projection operator projecting onto the
tangential plane of a level set of $\phi$.  Introducing the shorthand
$\phikalph = \partial_\alpha\phi$, we have $\nabla\phi = \Ehalph
\phikalph$ and
\begin{align}
  \label{eq:eval_diffproj}
  \nabla\cdot &\left(1-\normhat:\normhat\right) \nabla = \nonumber\\
& \quad\quad \nabla\cdot \bigg(\Ehmu \partial_{\mu} - \frac{1}{(\nabla\phi)^2}
\Ehalph \phikalph g^{\beta\mu} \phikbet \partial_\mu\bigg)  = \nonumber\\
& \quad\quad \frac{1}{\sqrt{g}} \partial_\nu \sqrt{g} g^{\nu\mu}
\left(\partial_{\mu}-\frac{1}{(\nabla\phi)^2}\phikmu g^{\alpha\beta} \phikalph
\partial_\beta\right)\>.
\end{align}
The third expression is obtained from the second applying the
divergence operator \eqref{eq:divergence} and renaming $\alpha\to\mu$,
$\beta\to\alpha$, $\mu\to\beta$ in the three pairs of ``mute''
indices.

To expand this operator in powers of $W$ in the inner domain, we
introduce an abbreviation for a normalized gradient of $\Phi$, being
of order $W^0$:
\begin{equation}
  \label{eq:nabphisq}
\begin{aligned}
  (\nabla\Phi)^2 &\equiv \frac{1}{W^2} (\tnabla\Phi)^2 \>, \\
  (\tnabla\Phi)^2 &=  W^2 \Phikalph g^{\alpha\beta} \Phikbet =
         \Phikrho^2 +  \frac{W^2}{g} \Phiks^2 \>.
\end{aligned}
\end{equation}
Inserting this into \eqref{eq:eval_diffproj} and carrying the
expansion to formal order $W^0$, we find first that the order
$W^{-2}$ terms (containing two derivatives with respect to
$\rho$) cancel each other.  The remainder reads
\begin{align}
 \label{eq:diffprojorder1}
 \nabla\cdot &\left(1-\normhat:\normhat\right) \nabla = \nonumber\\
&\frac{1}{\sqrt{g}}\,\partial_\rho \frac{1}{\sqrt{g}(\tnabla\Phi)^2}\left(
\Phiks^2 \partial_\rho - \Phikrho\Phiks \partial_\s\right)
 \nonumber\\
 &+ \frac{1}{\sqrt{g}}\,\partial_\s \frac{1}{\sqrt{g}}
 \left(\partial_\s
-\frac{\Phikrho \Phiks}{(\tnabla\Phi)^2}
 \partial_\rho\right) + O(W)\>,
\end{align}
and this expression still contains derivatives with respect to $\rho$.
However, {\em if} the leading-order solution $\Phinul$ depends on
$\rho$ only, as it did in the last section, then all the derivatives
of $\Phi$ with respect to $\s$ are $O(W)$ at least, and since
\hbox{$\sqrt{g} = 1+O(W)$}, Eq.~\eqref{eq:diffprojorder1} reduces to
$\nabla\cdot (1-\normhat:\normhat) \nabla = \partial_\s^2 + O(W)$,
i.e., at leading order the operator indeed becomes the
Laplace-Beltrami operator on the surface.

This then suggests to replace the phase-field equation
\eqref{eq:phidyn_c} with
\begin{equation}
  \label{eq:phidyn_c_tens}
  \frac{\partial \phi}{\partial t} = M
\nabla \cdot \left(1-\normhat:\normhat\right) \nabla \,
\frac{1}{W^2}\delta\mutil(\nabla^2\phi,\phi) \>,
\end{equation}
where $\delta\mutil$ is unchanged from \eqref{eq:phidyn_c} but
$M$ is a constant mobility now.

In this model, the equation for the velocity would appear at the
next-to leading order already and take the form
\begin{eqnarray}
\label{eq:app_veloc}
v_n \partial_\rho \Phinul&=&
M \partial_{\s\s}
\bigg\{\partial_{\rho\rho} \Phione
 + \kappa \partial_\rho \Phinul- 2 f''(\Phinul)\,\Phione\bigg\}
 \>. \nonumber\\
\end{eqnarray}
Because the operators $\linopl$ [defined in Eq.~(\ref{eq:linop})] and
$\partial_{\s\s}$ commute, Fredholm's alternative is applicable the
same way as in the nonconservative case.  This eliminates $\Phione$
from the equation and produces the correct sharp-interface limit.

In spite of this enjoyable state of affairs, model
\eqref{eq:phidyn_c_tens} fails much more miserably than
\eqref{eq:phidyn_c}. The reason is that the zeroth-order solution is
not unique.  In fact, the leading-order outer equation
\begin{equation}
\label{eq:outer0}
\nabla \cdot \left(1-\normhat^{(0)}:\normhat^{(0)}\right)
 \nabla f'(\phinul) = 0
\end{equation}
is solved by {\em any} (differentiable) function $\phinul$
satisfying the boundary conditions: obviously we have $\nabla
f'(\phinul) = f''(\phinul)\nabla\phinul$, whence
\begin{align}
  \label{eq:effect_proj}
  \normhat^{(0)}:\normhat^{(0)} \nabla f'(\phinul) &= -\normhat^{(0)}
f''(\phinul) \abs{\nabla\phinul} \nonumber\\
&= f''(\phinul) \nabla\phinul = \nabla f'(\phinul)\>,
\end{align}
which implies $(1-\normhat^{(0)}:\normhat^{(0)}) \nabla f'(\phinul) =
0$ for all functions $f'$ of $\phinul$. Intuitively, this behavior can
be easily understood for a planar interface. Then the equation of
motion \eqref{eq:phidyn_c_tens} strictly contains only derivatives of
the phase field parallel to the interface, and the profile in the
perpendicular direction therefore remains completely undetermined.

It can be said that this model fails for reasons complementary to
those of the scalar model.  Whereas we had one equation too many in
that case, adding a constraint to the desired sharp-interface
dynamics, now we have one equation too few, as there is nothing in the
model fixing $\phinul$.  If we had the right $\phinul$, the tensorial
model would work perfectly.

\section{Modified tensorial mobility models with correct asymptotics
\label{sec:correct}}

In order to obtain a model not plagued by either of the disadvantages
of the two cases discussed, it appears that it is useful to combine ideas
from both.  While it is certainly desirable to have a differential
operator that itself approaches the surface Laplacian, it should do so
only for phase field functions that have the correct leading-order
profile.

\subsection{Locally conservative model \label{sec:fct}}

One way to achieve this goal is to modify $1-\normhat:\normhat$ into
\begin{equation}
\label{eq:defQ}
Q\equiv 1-W^2 \frac{\nabla\phi:\nabla\phi}{4 f(\phi)}
= 1 -  \frac{W^2 (\nabla\phi)^2}{4 f(\phi)} \normhat:\normhat\>.
\end{equation}

If we replace the projection operator in Eq.~\eqref{eq:phidyn_c_tens}
by $Q$, then the {\em outer} equation at leading order will have the
same differential operator as the scalar model with constant $M$.

On the other hand, in the {\em inner} domain, we have, {\em provided}
$\Phinul$ solves the differential equation \eqref{eq:diffeq_Phi0},
$W^2(\nabla\Phi)^2 =\Phikrho^2 +O(W^2)=4
f(\Phi)+O(W)$ [this follows from Eqs.~\eqref{eq:nabphisq},
\eqref{eq:diffeq_Phi0_firstint}, and \eqref{eq:fphi}], whence
$Q\approx 1-\normhat:\normhat$.

This approximation is accurate up to $O(W)$ only, which is not
sufficient, because the order $W$ correction would enter as a
bothersome additive term in the next-to leading order inner equation.

A better inner approximation to $1-\normhat:\normhat$ than just $Q$ is
provided by a minor modification.  Obviously, we have
$Q=1-\normhat:\normhat+O(W)\,\normhat:\normhat$ in the inner
region.  Taking this to some integer power $m$ we get, because
$1-\normhat:\normhat$ and $\normhat:\normhat$ are orthogonal
projectors:
\begin{equation}
\label{eq:Qm}
Q^m =  1-\normhat:\normhat+O(W^m)\,\normhat:\normhat \>.
\end{equation}

These considerations lead us to make the ansatz
\begin{eqnarray}
  \label{eq:phidyn_cval}
  \frac{\partial \phi}{\partial t} &=&
 M \nabla \cdot \jvec \nonumber \\
\jvec &=& Q^m \nabla  \frac{1}{W^2}\,
\delta\mutil(\nabla^2\phi,\phi) \>,\\
\delta\mutil &\equiv&
-W^2\nabla^2 \phi + 2 f'(\phi)
\nonumber
\end{eqnarray}
and leave the precise choice of the value of
$m$ for later -- it will be suggested by the asymptotic analysis.

The corresponding inner equations are
\begin{eqnarray}
  \label{eq:phidyn_c_innval}
  \partial_t \Phi -\frac{1}{W} v_n \partial_\rho \Phi &=&
M \nabla \cdot Q^m \nabla \frac{1}{W^2}\,
\delta\mutil(\nabla^2\Phi,\Phi)\>,\nonumber\\
 \delta\mutil(\nabla^2\Phi,\Phi) &=&
-\frac{1}{\sqrt{g}}\,\partial_\rho \sqrt{g}\,
 \Phikrho \\
 && - W^2\frac{1}{\sqrt{g}}\,\partial_\s \frac{1}{\sqrt{g}}
 \Phiks
+ 2 f'(\Phi) 
\>,\nonumber
\end{eqnarray}
with
\begin{align}
  \label{eq:diffop2_inn}
\nabla \cdot Q^m \nabla = &\frac{1}{\sqrt{g}} \partial_\nu \sqrt{g} g^{\nu\mu}
\bigg\{\partial_\mu-\bigg[1-\Big(1-\frac{W^2
(\nabla\Phi)^2}{4 f(\Phi)}\Big)^m\bigg] \nonumber\\
&\times\frac{1}{(\nabla\Phi)^2}\Phikmu g^{\alpha\beta} \Phikalph
\partial_\beta\bigg\}
\>.
\end{align}

\subsubsection{Leading order}
In the outer equations, $Q$ becomes the identity operator to leading
order, i.e., $Q^{(0)}(\phinul) = 1$, and at the lowest order in
$W$, we have
\begin{equation}
  \label{eq:valout}
  \nabla^2 f'(\phinul) = 0 \>,
\end{equation}
a Laplace equation
that we know to be uniquely solvable for $f'(\phinul)$ with Dirichlet
boundary conditions at infinity.  This boundary condition is even
homogeneous (except possibly in a part of the boundary region at
infinity having a size of order $W$), leading to the unique solution
$f'(\phinul)\equiv 0$.  This leaves us with the three possibilities
$\phinul=0$, $\frac12$, $1$, of which $\phinul=0$ or $\phinul=1$
are realized, according to the particular boundary condition on
$\phinul$.

Again, $\phi=0$ and $\phi=1$ are solutions to the outer problem at all
orders of $W$.  Admittedly, the operator $Q$ becomes indefinite
at order $W^2$ for $\phi=0$ and $\phi=1$ [because of the
denominator $f(\phi)$], but this does not matter, since the
expression for $\delta \mutil$ alone is zero already at $\phi=0$ and
$\phi=1$.

The leading-order inner equation reads [$g=1+O(W)$]
\begin{eqnarray}
  \label{eq:phydin_cval_innlead}
  \partial_\rho  \left[ 1-\frac{\left(\Phikrhonul\right)^2}{4 f(\Phinul)}
\right]^m \!\! \partial_\rho
\bigg(\partial_{\rho\rho} \Phinul -2 f'(\Phinul)\bigg) = 0\>.\nonumber\\
\end{eqnarray}
Clearly, this is solved by $\Phinul=\frac12 \,(1-\tanh\rho)$,
which makes both the expression in brackets and in large parentheses
vanish.  If we require $m$ to be even, this solution is moreover
unique (up to translations, which are eliminated by the requirement
that the interface be at $\rho=0$).  For as soon as we assume
$(\Phikrhonul)^2 \ne 4 f(\Phinul)$, the $m$th power of the bracket
expression will be positive, allowing us to use similar arguments as
in Sec.~\ref{sec:expansions} between Eqs.~\eqref{eq:phydin_c_innlead}
and \eqref{eq:phidyn_c_innlead_res} to prove that $\partial_{\rho\rho}
\Phinul -2 f'(\Phinul)=0$, and hence the bracket expression must be
zero, contrary to our assumption.  Thus we do get a definite solution
for $\Phinul$ from the inner equation, which moreover shows that at
leading order of the inner expansion the second-order $\rho$
derivatives of the operator $\nabla\cdot Q^m \nabla$ cancel each
other.

\subsubsection{Next-to-leading order}

To simplify computations at the next order, we first expand
$\nabla\cdot Q^m\nabla$ up to formal order $W^0$.  This produces
\begin{align}
  \label{eq:expansion_nqmn}
\nabla&\cdot Q^m \nabla =  \nonumber\\
&\phantom{+}\frac{1}{\sqrt{g}}\,\partial_\s \frac{1}{\sqrt{g}}\,\partial_\s
+\frac{1}{\sqrt{g}}\, \frac{1}{W^2}\, \partial_\rho \sqrt{g}
\left(1-\frac{W^2 (\nabla\Phi)^2}{4 f(\Phi)}\right)^m \partial_\rho
\nonumber\\
&-\frac{1}{\sqrt{g}}\,\partial_\rho  \frac{1}{\sqrt{g}} \frac{\Phiks}{\Phikrho}
\partial_\s
-\frac{1}{\sqrt{g}}\,\partial_\s\frac{1}{\sqrt{g}}  \frac{\Phiks}{\Phikrho}
 \partial_\rho
\nonumber\\
&+\frac{1}{\sqrt{g}}\,\partial_\rho \frac{1}{\sqrt{g}} \,\frac{\Phiks^2}{\Phikrho^2}
 \partial_\rho
 + O(W)\>,
\end{align}
Given that $\Phinul$ is a function of $\rho$ only, we realize that the
third and fourth terms on the right hand side are $O(W)$,
containing derivatives with respect to $\s$  of $\Phi$, the
fifth is even $O(W^2)$, so these terms may be dropped
immediately in an expansion up to $O(1)$.  The second term on the
right hand side owes its existence to the fact that $Q$ is not exactly the
projection operator on $\normhat$ [note that no such term is present
in Eq.~\eqref{eq:diffprojorder1}] and it has a prefactor of
$1/W^2$ due to the double derivative in $\rho$.  This term
which is desirable at leading order, because without it we would not
have a determinate zeroth order solution $\Phinul$, is somewhat
disturbing at the next order.  Since the order of this term is
$O(W^{m-2})$, we can make it small by choosing $m\ge 3$, i.e.,
restricting ourselves to even $m$ for the reasons discussed before, we
set $m=4$.  Then the only remaining term on the right hand side of
Eq.~\eqref{eq:expansion_nqmn} up to order $W^0$ is the first
term, which is the desired surface Laplacian.

Using this result, we can write the next-to-leading (nontrivial) order
inner equation
\begin{equation}
  \label{eq:nextorder}
\begin{aligned}
 - v_n \partial_\rho \Phinul&= M \partial_{\s\s}
 \delta\mutil^{(1)}\>,
\\
\delta\mutil^{(1)}&=
-\linopl \Phione
 - \kappa \partial_\rho \Phinul
 \>,
\end{aligned}
\end{equation}
again with $\linopl$ as given in Eq.~\eqref{eq:linop}.

Note that we actually seem to have skipped orders here.  The
leading-order inner equation is formally $O(W^{-4})$, but once
the zeroth-order inner solution is fixed, the differential operator
$\nabla\cdot Q^m \nabla$ is, according to \eqref{eq:expansion_nqmn},
of order $W^{\max(0,m-2)}$ only, so the order $W^{-3}$ vanishes
identically.  The order $W^{-2}$ is satisfied automatically,
because the zeroth-order chemical potential is zero; the next
nontrivial order is $W^{-1}$.  Alternatively, one may say that
the effective leading order has become $W^{-2}$.

The total linear operator in front of $\Phione$ becomes
$-\partial_{\s\s}\linopl$.  It is hermitian, because its hermitian
factors commute.  Hence, $\partial_\rho\Phinul$ is a left null
eigenfunction.  Multiplying \eqref{eq:nextorder} from the left by it,
integrating with respect to $\rho$ from $-\infty$ to $\infty$, we
obtain Eq.~\eqref{eq:veloc_scalmod}.
This proves that the considered phase-field model based on a
modified tensorial mobility has the correct asymptotic behavior for
small $W$, neither overconstraining the system by adding, nor leaving
it indeterminate by losing equations.

Clearly, Eq.~\eqref{eq:phidyn_cval} establishes a \emph{local}
conservation law for $\phi$, i.e., the rate of change of the integral
of $\phi$ over some control volume is given by the integral of the
current $\jvec$ associated with $\phi$ over the surface of the volume,
and this holds for \emph{arbitrarily small} volumes. $\phi$ is the
density of a conserved quantity.  In particular, for a system with
boundaries through which there is no flux, the volume integral of
$\phi$ will be conserved. Therefore, we will denote the model
discussed in this section as the \emph{locally conservative tensorial}
or \emph{LCT model}.

\subsection{Globally conservative model \label{sec:act}}
If one is willing to give up the conservative nature of the
phase-field equations themselves and requires the conservation law
only in the asymptotic limit, an even simpler construction is feasible.

Consider the model
\begin{equation}
\label{eq:spatschek_model}
\begin{aligned}
\frac{\partial \phi}{\partial t} &=
M \nabla\cdot P\nabla\frac{1}{W^2}\delta\mutil +
N \left[\left(\normhat\cdot\nabla\right)^2\phi-\frac{2}{W^2}f'(\phi)\right] \\
P&=1-\normhat:\normhat\\
\delta\mutil&=-W^2 \nabla^2 \phi+ 2 f'(\phi)
\end{aligned}
\end{equation}
with both $M$ and $N$ positive constants.  With $N=0$, this reduces to
the tensorial model of Sec.~\ref{sec:tensorial}. With \hbox{$N>0$, it}
can be shown along similar lines as in Sec.~\ref{sec:expleading} that
the leading-order outer solution for $\delta\mutil$ is unique leading
to the solutions $\phi=0$ and $\phi=1$, depending on the boundary
conditions at infinity.  Moreover, the leading-order inner solution
with boundary conditions $\lim_{\rho\to-\infty}\Phinul= 0$,
$\lim_{\rho\to\infty}\Phinul= 1$ can be shown to be unique up to
translations along $\rho$ and is given by \eqref{eq:solution_Phi0}
after requiring $\rho=0$ to correspond to the value $\frac12$ of the
phase field.

The role of the $N$ term is only to fix the profile of the phase field
at leading order, otherwise it is constructed so as to not affect the
normal velocity of the interface.  Once $\Phinul$ is set, the
next-to-leading order inner equation reads:
\begin{equation}
\label{eq:spatschek_inn}
\left(M \partial_{ss}-N\right) \linopl \Phione
= v_n \partial_\rho \Phinul - M \partial_{ss}\kappa \partial_\rho \Phinul\>,
\end{equation}
and since $M \partial_{ss}-N$ commutes with $\linopl$, we obtain the
desired sharp-interface limit again. Our numerical investigations
indeed show that the results depend only weakly on the choice of the
parameter $N$, even for moderate separation of the length scales. We
find $N\le 2.5 M/W^2$ 
to already give satisfactory results -- there are only small
differences to results obtained when $N$ is two orders of magnitude
smaller, i.e., for $N = 1.25\times 10^{-2} M/W^2$.

While the model \eqref{eq:spatschek_model} is asymptotically
conservative, it is desirable to have exact global conservation of the
phase-field, because this will render long-time simulations more
reliable. As the model stands, one might be obliged to choose the
interface width smaller as the simulation time becomes larger, which
is certainly something one would wish to avoid. Therefore, even though
the violation of phase-field conservation is small and the model would
already be useful in its present form, let us look for an improvement
restoring global conservation. By this we mean that $\phi$ need not be
the density of a conserved quantity, hence its time derivative need
not be the divergence of a current, but for no-flux boundary
conditions, the total volume integral of $\phi$ should remain
conserved.  This can of course be achieved via the introduction of a
Lagrange parameter:
\begin{equation}
\label{eq:spatschek_model_lagrange}
\frac{\partial \phi}{\partial t} =
M \nabla\cdot P\nabla\frac{1}{W^2}\delta\mutil +
N \left[\left(\normhat\cdot\nabla\right)^2\phi-\frac{2}{W^2}f'(\phi)\right]
- \la(\rvec, t)\>.
\end{equation}

Here, we have allowed $\la$ to depend on $\rvec$ which gives
useful additional freedom for improvement of the model as we shall see
immediately.  If $\la$ were restricted to being a simple number,
it would have to to have the value
\newcommand{\phiav}{\bar{\phi}}
\begin{eqnarray}
\label{eq:simplelambda}
\la &=& \frac{1}{V} \int_V \,dV\>
\frac{\partial\phi_{\text{old}}}{\partial t} \nonumber\\
               &=&  \frac{N}{V} \int_V \,dV\>
 \left[\left(\normhat\cdot\nabla\right)^2\phi-\frac{2}{W^2}f'(\phi)\right] \>,
\end{eqnarray}
where $\partial\phi_{\text{old}}/{\partial t}$ is the time derivative
of the phase field according to \eqref{eq:spatschek_model} and $V$ is
the volume (or area, in 2D) of the system.  Since the first term of
the right hand side of \eqref{eq:spatschek_model_lagrange} is
conservative anyway, it drops out of the calculation of $\la$, if no
fluxes through the system boundary are present.  A drawback of the
formulation \eqref{eq:simplelambda} is that it would lead to a
modification of the phase field in the bulk from the equilibrium
values 0 and 1, as soon as the Lagrange multiplier became nonzero.
This can be avoided by taking advantage of the liberty to make $\la$
vary in space (i.e., we consider a whole set of Lagrange multipliers,
not just one).  If we take $\la$ of the form
\begin{equation}
\label{eq:lambdart}
\la(\rvec, t) = \frac{|\nabla \phi|}{\int_V \,dV\> |\nabla \phi|}  \int_V \,dV\>
 \left[\left(\normhat\cdot\nabla\right)^2\phi-\frac{2}{W^2}f'(\phi)\right]\>,
\end{equation}
the global conservation law is restored without any modification of
the bulk solutions. We will call the model described by
Eqs.~\eqref{eq:spatschek_model_lagrange} and \eqref{eq:lambdart} the
\emph{globally conservative tensorial} or \emph{GCT model}.

Of course, we have to verify that the introduction of the Lagrange
parameter does not destroy the asymptotic validity of the model.
Clearly, the parameter disappears from the leading order of the
equation; but the interface velocity is determined at next-to leading
order, and in general one would expect $\la(\rvec, t)$ to contribute
to the equation at that order. This turns out not to be the case
and is due to the judicious choice of the form of the parameter, as we
shall see now.

The next-to leading order inner equation can be written
\begin{align}
\label{eq:spatschek_laminn}
&\left(M \partial_{ss} -N \right) \linopl \Phione
+ N \frac{\partial_\rho \Phinul}{\int dV \partial_\rho \Phinul}
 \int dV \linopl \Phione \nonumber\\
&\hspace*{2cm}
= v_n \partial_\rho \Phinul - M \partial_{ss}\kappa \partial_\rho \Phinul\>,
\end{align}
and we are in the awkward situation that the linear operator acting on
$\Phione$ is not self-adjoint, so the application of Fredholm's
alternative becomes nontrivial.  However, the equation contains several
terms $\propto \partial_\rho \Phinul$, which suggests to have
$\linopl$ act on it, leading first to the much simpler equation
\begin{equation}
\label{eq:laminn_simple}
\linopl \left(N-M \partial_{ss}  \right) \linopl \Phione =0 \>,
\end{equation}
because $\linopl \partial_\rho\Phinul =0$.  But here the operator acting on
$\Phione$ is semipositive, the operator sandwiched by the two
$\linopl$s strictly positive.  Hence, we may conclude that
$\linopl\Phione =0$.  But then the left-hand side of
\eqref{eq:spatschek_laminn} is zero, meaning that the linear equation
is in fact homogeneous and the right-hand side has to vanish, too.
This implies
\begin{equation}
\label{eq:spatschek_lam_vn}
 v_n =  M \partial_{ss}\kappa\>,
\end{equation}
the sought-for asymptotic result for the interface velocity.  It also
implies that the Lagrange multiplier is $O(1)$, instead of $O(W^{-1})$,
i.e., it is by a factor of the order of $(W\kappa)^2$ smaller than the
leading-order terms of the equation.

This supports what we can point out on the basis of numerical studies:
for reasonable separation of length scales as they appear in typical
simulations, the influence of the Lagrange parameter is negligibly
small, at least for not too long time scales.

\section{Analytic linear stability analysis \label{sec:linstab}}


A linear stability analysis of a stationary planar front may be useful
in trying to differentiate between the models.  Clearly, one should
require that the spectrum obtained for the phase-field model reduces
to that of the sharp-interface model [e.g., Eqs.~\eqref{eq:disprel_nc}
or \eqref{eq:disprel_co}] in the limit of small interface width.

For simplicity, let us first consider the nonconservative model described
by \eqref{eq:phidyn_nc}. A planar front solution is given by
\begin{equation}
\label{eq:planar_front}
\phi = \phizero(z) = \Phizero(Z) = \frac12 \left(1-\tanh Z\right)\>,
\qquad Z=\frac{z}{W}
\end{equation}
Adding a small perturbation $\delta \phi$, i.e., setting
$\phi=\phizero+\delta \phi(x,z,t)$, we obtain the linearized equation
\begin{equation}
  \label{eq:phidyn_nc_lin}
 \frac{\partial \delta \phi}{\partial t} =
M \biggl( \nabla ^2
-  \frac{2}{W^2} f''(\phizero)
\biggr) \, \delta\phi \>.
\end{equation}
Using the ansatz
\begin{equation}
\label{eq:pert_ansatz}
\delta \phi(x,z,t) = \Psi(Z) e^{\im k x + \omega t}
\end{equation}
we obtain the eigenvalue problem
\begin{equation}
\label{eq:eigenval_nc}
\omega\Psi = \frac{M}{W^2}\left(\partial_{ZZ}-W^2 k^2
 -2 f''(\Phizero)\right) \Psi\>.
\end{equation}
While this is a linear problem, it is one that does not have constant
coefficients, so an exact solution is not readily available.  Instead,
we must rely on asymptotic analysis again to make progress.  However,
as it turns out by reinserting the found eigenfunctions and
eigenvalues into \eqref{eq:eigenval_nc}, the expansion provides exact
results in this case.  Details of the calculation are given in
App.~\ref{sec:linstab_det}. We find two branches of the spectrum with
\begin{equation}
\label{eq:spectr_nc}
\begin{aligned}
\omega_a &= -\frac{4M}{W^2}-M k^2 \>,\\
\omega_b &= -M k^2 \>.
\end{aligned}
\end{equation}
For the first branch, the eigenfunction does not vanish as
$Z\to\pm\infty$, for the second it does and moreover is proportional
to $\Phizero'(Z)$. As $Wk \ll 1$, any contribution of the perturbation
containing the first eigenfunction will decay fast, leaving a
remainder that decays with rate $\omega_b$, which corresponds to the
dispersion relation of the sharp interface-limit,
Eq.~\eqref{eq:disprel_nc}.  Note also that if we assume our initial
perturbation to describe a slightly perturbed planar front,i.e.,
\begin{equation}
\label{eq:planar_front_pert}
\phi  = \frac12 \Bigl\{1-\tanh\left[Z-\delta\zeta(x)\right]\Bigr\}\>,
\end{equation}
then we have
\begin{equation}
\delta\phi=-\frac{\partial\Phizero}{\partial Z}\delta\zeta\>,
\end{equation}
hence the perturbation \emph{is} $\propto \Phizero'(Z)$, so the
relevance of the second eigenvalue is obvious.

With these preliminaries, the linear stability analysis of the LCT
model becomes more or less trivial. Linearizing the equation of motion
\eqref{eq:phidyn_cval} about $\phizero$, we obtain
\begin{equation}
  \label{eq:phidyn_co_lin}
  \frac{\partial\delta \phi}{\partial t} =
  M \nabla Q_0^4 \nabla  \left(-\nabla^2  +
 \frac{2}{W^2} f''(\Phizero)\right)\delta\phi\>,
\end{equation}
where
\begin{equation}
 Q_0 = Q(\Phizero) = 1-\ez:\ez\>,
\end{equation}
as the level set normals are in the $z$ direction and $W^2
\left(\nabla\Phizero\right)^2/4 f(\Phizero)=1$. But then $Q_0^4=Q_0$,
and $\nabla Q_0^4 \nabla=\partial_x^2$.  Hence,
\begin{equation}
  \label{eq:phidyn_co_lin2}
 \frac{\partial \delta \phi}{\partial t} =
M \partial_{xx}\biggl( \nabla ^2
-  \frac{2}{W^2} f''(\phizero)
\biggr)\, \delta\phi
\end{equation}
and after inserting ansatz \eqref{eq:pert_ansatz}, one gets
\begin{equation}
\label{eq:eigenval_fct}
\omega\Psi = \frac{M k^2}{W^2}\left(\partial_{ZZ}-W^2 k^2
 -2 f''(\Phizero)\right) \Psi\>.
\end{equation}
But this is the same eigenvalue problem as \eqref{eq:eigenval_nc} with
$M$ replaced by $Mk^2$. So we obtain a two-branch dispersion relation
again, this time with eigenvalues
\begin{equation}
\begin{aligned}
\omega_a &= -\frac{4M k^2}{W^2}-M k^4 \>,\\
\omega_b &= -M k^4 \>,
\end{aligned}
\end{equation}
which in the light of the preceding discussion is a reasonable result
(the eigenvalue with large absolute value is negative and influences
the dynamics for a short time only).

Analysing the GCT model is only slightly more involved, at least in
the form without the Lagrangian multiplier. One first derives that,
for a perturbation about $\Phizero$ the simplification
\begin{equation} \delta
\left(\normhat\nabla\right)^2\phi =
\frac1{W^2}\partial_{ZZ}\delta\phi
\end{equation}
holds true. Then the
linearization of Eq.~\eqref{eq:spatschek_model}  reads
\begin{equation}
\label{eq:spatschek_model_lin}
\begin{aligned}
\frac{\partial \delta\phi}{\partial t} &=
M \partial_{xx}\biggl(-\nabla^2+\frac{2}{W^2} f''(\Phizero)\biggr) \,\delta\phi \\&\quad+
\frac{N}{W^2} \left(\partial_{ZZ}-f''(\Phizero)\right) \delta\phi \>.
\end{aligned}
\end{equation}
After using the ansatz \eqref{eq:pert_ansatz}, one can cast the
resulting eigenvalue problem into the form (see App.~\ref{sec:linstab_det})
\begin{equation}
\label{eq:eigenval_act}
\left(\omega-Nk^2\right)\Psi = \frac{M k^2+N}{W^2}\left(\partial_{ZZ}-W^2 k^2
 -2 f''(\Phizero)\right) \Psi\>.
\end{equation}
This again has the same form as \eqref{eq:eigenval_nc}, now with
$M$ replaced by $Mk^2+N$ and $\omega$ replaced by $\omega-Nk^2$, leading to
\begin{equation}
\begin{aligned}
\omega_a &= -\frac{4(M k^2+N)}{W^2}-M k^4 \>,\\
\omega_b &= -M k^4 \>.
\end{aligned}
\end{equation}
Considering now the form \emph{with} the Lagrangian multiplier,
\eqref{eq:spatschek_model_lagrange} with \eqref{eq:lambdart}, we note
that
\begin{equation}
\delta\la = \frac{\abs{\nabla\Phizero}}{\int_V d^3x \abs{\nabla\Phizero}}
\int_V d^3x \frac{1}{W^2}\left(\partial_{ZZ}-2 f''(\Phizero)\right) \delta\phi\>,
\end{equation}
which becomes zero, if we insert the eigenfunction $\Phizero'(Z)$
belonging to $\omega_b$ (see App.~\ref{sec:linstab_det}).  So this
eigenvalue, which is the relevant one, remains unchanged.  We can no
longer exactly calculate the other eigenvalue nor the corresponding
eigenfunction, but we anticipate that this eigenvalue still behaves as
$1/W^2$ at leading order and is negative, so it does not dominate the
asymptotic behavior.

Unfortunately, an analysis of the SM model along the same lines turns
out to be much more complicated. This may be traced back to the fact
that the mobility function vanishes in the outer domain, leaving no
useful linearized outer equation. In fact, the outer equation formally
becomes $\partial_t \delta\phi = 0$, saying that at linear order
perturbations will not decay at infinity.  In reality, this means that
linearization is not legitimate, as perturbations will decay via
nonlinear relaxation -- the leading-order nonlinearity is larger than
the (vanishing) linear expression. An analysis of the linearized inner
equation with the requirement that $\delta\Phi\to 0$ for
$Z\to\pm\infty$ does not produce any solutions for eigenvalues assumed
to diverge more slowly as $1/W^4$ (or not at all).  If the requirement
$\delta\Phi\to 0$ is given up, one may construct perturbation
eigenfunctions, but these contain polynomially diverging terms at
infinity. On the other hand, eigenvalues of the form $\omega =
\omega_{-4}/W^4+\dots$ could not be investigated by asymptotic
analysis, because with this assumption the lowest-order perturbative
problem remains unsolvable analytically. Note, however, that one
eigenvalue of the form $\omega =-M k^4+O(W)$ has to be expected for
any phase-field model trying to meaningfully approximate the
sharp-interface dynamics \eqref{eq:vn_surfdiff}.

It is tempting to speculate that the fact of an undesirable
restriction arising from the asymptotic analysis of the SM model and
the unfeasibility of asymptotic analysis in a linear stability
calculation of the model have to do with each other. This fits nicely
with the observation that the linearized RRV model does have a usable
outer equation. While its mobility function $B(\phi)$ is zero in
the bulk as well, the presence of the stabilizing function $g(\phi)$
prevents the reduction to a static result. Writing the model in one equation
\begin{equation}
  \label{eq:phidyn_c_voigt_sh}
  \frac{\partial \phi}{\partial t} =  M \nabla \cdot
 B(\phi) \nabla \frac{1}{g(\phi)}\left(
- \nabla^2 \phi+ \frac{2}{W^2}f'(\phi)\right)\,,
\end{equation}
we can recast the differential operator in front of the parentheses
\begin{align}
\nabla \cdot B(\phi) \nabla \frac{1}{g(\phi)} &= \nabla^2 \frac{B(\phi)}{g(\phi)}
-\nabla\cdot\left(\nabla B(\phi)\right) \frac{1}{g(\phi)} \nonumber\\
&= \frac65 \nabla^2-\frac{12}{5} \nabla\cdot(1-2\phi)\frac{\nabla\phi}{\phi(1-\phi)}
 \>,
\end{align}
and the last term remains regular.
The linearized equation of motion then becomes
\begin{equation}
\label{eq:voigt_lin}
\begin{aligned}
 \frac{\partial \delta \phi}{\partial t} =
& \frac65 M \left[\nabla^2+\frac{4}{W^2}\partial_Z(1-2\Phizero)\right]
\left(- \nabla^2+ \frac{2}{W^2}f''(\Phizero)\right)\delta\phi\>,
\end{aligned}
\end{equation}
which for $\abs{Z}\gg1$ turns into
\begin{equation}
\label{eq:voigt_lin_out}
\begin{aligned}
 \frac{\partial \delta \phi}{\partial t} =
& \frac65 M \left[\nabla^2+\frac{4}{W}\text{sign} z \partial_z\right]
\left(- \nabla^2+ \frac{4}{W^2}\right)\delta\phi\>,
\end{aligned}
\end{equation}
a perfectly sensible outer equation.

Because the inhomogeneus fourth-order linear differential equations
resulting from \eqref{eq:voigt_lin} at successive steps of the
expansion are difficult to solve (this is in part due to the linear
operator being nonhermitian) and the impossibility of getting analytic
results for the SM model necessitates a numerical investigation
anyway, we will not pursue the analytic discussion of the RRV model
any further. We have checked that $\Phizero^{\prime}(Z) e^{\im
  kx+\omega t}$ is not an exact solution to the linearized equation
here, so it appears quite likely that the relevant branch of the
dispersion relation contains $W$ dependent corrections, i.e. $\omega =
-M k^4 + a W^2$.

\section{Numerical simulations\label{sec:numerical}}

To determine growth rates numerically, we simulate the temporal
evolution of a planar front subject to a sinusoidal perturbation. This
allows the empirical determination of the main branch of the
dispersion relations of the models considered, i.e., of the largest
eigenvalue ($\omega_b$). Moreover, to assess the behavior of the
models in a growth situation, we include the coupling to elastic
fields, i.e., we simulate the Grinfeld instability. This is easily
achieved by replacing the ``free'' chemical potential $\delta\mutil$
with one that includes the correct elastic energy contribution.

In particular, we set
\begin{eqnarray}
  \label{eq:phidyn_c_gf}
\delta\mutil(\nabla^2\phi,\phi) &\equiv&
-W^2\nabla^2 \phi + 2 f'(\phi) + \frac{W}{3\gamma} h'(\phi) \potelast
\>,
\end{eqnarray}
with
\begin{equation}
  \label{eq:defpotelast}
\potelast = \left(\mus-\mustil\right) \sum_{i,j} \ud_{ij}^2 +
\frac{\lambda-\lamtil}{2} \Big(\sum_{k}\ud_{kk}\Big)^2 \>,
\end{equation}
where $\mus = E/[2(1+\nu)]$ is the shear modulus or first Lam\'e
constant, $\lambda = E\nu/[(1+\nu)(1-2\nu)]$ the second Lam\'e
constant in the solid phase and $\mustil$ and $\lamtil$ are the
corresponding quantities in the second phase. If the second phase is a
liquid, $\mustil=0$, if it is vacuum, $\mustil=\lamtil=0$. From now
on, we will focus on the vacuum case, as it is particularly
interesting for diffusion along a free surface. Equation
\eqref{eq:phidyn_c_gf} was originally introduced as a phase-field
model for two coherent solid phases in contact with each other in
\cite{spatschek07}. Moreover, it has been discussed in
\cite{kassner01} that modeling a liquid (or vacuum) as a shear-free
solid faithfully captures the physics of the system in spite of the
property of formally coherent strains.  This is essentially due to the
fact that only the divergence of the strain tensor is a physically
meaningful quantity in the shear-free phase.

By $u_i$, we denote the displacement field. In addition, equations for
the strains $u_{ij} = \frac{1}{2}\left(\partial \ud_i/\partial x_j
  +\partial \ud_j/\partial x_i\right)$ have to be provided, which are
given by the mechanical equilibrium conditions for the generalized
stress tensor $\ssigma_{ij}$
\begin{equation}
  \label{eq:elasteqs_ph}
  \sum_j \frac{\partial\ssigma_{ij}}{\partial x_j} = 0 \>, \qquad
  \ssigma_{ij} = h(\phi) \,\sigma_{ij}  \>.
\end{equation}
The function $h(\phi)$ is defined in \eqref{eq:hphi} and $\sigma_{ij}$ is
related to $u_{ij}$ via Hooke's law
\begin{equation}
\sigma_{ij} = \frac{E}{1+\nu}\bigg(\ud_{ij}
+\frac{\nu}{1-2\nu} \sum_k \ud_{kk} \,\delta_{ij} \bigg)
\>.
\label{eq:hooke}
\end{equation}

That this model produces the right coupling to elasticity in the
sharp-interface limit has been shown in various places, e.g. in
\cite{kassner01} und \cite{kassner_condmat06}. A similar modification
of $\delta\muhat$ leads to the fully coupled RRV model \cite{raetz06}.

To simulate the Grinfeld instability in a strip geometry, a
dimensionless driving force $\driv$ is defined as
\begin{equation}
\driv = \frac{\delta^2 (\lambda + 2\mus)}{4\gamma L}\>,
\end{equation}
with $\delta$ being a fixed displacement by which the strip is
elongated in the direction parallel to the interface. We use a square
system of length $L$ and uniform grid spacing $\Delta x$. The
interface width, a purely numerical parameter, is chosen to be $W=5
\Delta x$. For a sinusoidal perturbation of a uniaxially strained
surface by $\delta y(x)=A_0 \sin(kx)$ we use a fixed wavenumber $k L=4
\pi$ and a small amplitude $A_0 k= \pi/20$. To obtain a good
separation of the characteristic wavelength of the pattern and the
interface width, we use $k W = 0.16$.  The imposed uniaxial stress is
given in the figure caption for each case.  After having determined
the maximum admissible time step, as discussed below, we take as
simulation time step $\Delta t = 5 \cdot 10^{-4} (\Delta x)^4/M$ for
the scalar models and $\Delta t = 5 \cdot 10^{-3} (\Delta x)^4/M$ for
the tensorial models. The Poisson ratio is chosen to be $\nu =
\lambda/[2(\lambda + G)] = 1/3$. For the GCT model, we use $N=1.25
M/W^2$, and for the SM model the standard choice $\tilde{M}(\phi) =
36M \phi^2(1-\phi)^2$.  To minimize the influence of the boundaries,
we use helical boundary conditions in the $x$ direction for the
displacement fields, i.e.  $u_x(L, y)=u_x(0, y)+ \delta$, $u_y(L,
y)=u_y(0, y)$.


Introducing the Griffith length $L_G$
\begin{equation}
        L_G = \frac{L(1-\nu)^2}{8 \driv (1-2\nu)}\>,
\end{equation}
we can write the dispersion relation, i.e., the spectrum of the linear
stability operator, as follows
\begin{equation}
\label{eq:disprel_gf}
\omega = M\left(\frac{k^3}{L_G}-k^4\right)\>,
\end{equation}
meaning that we have unstable modes at small $k$ ($k<1/L_G$) and
stable ones at large $k$ ($k>1/L_G$). To obtain the spectrum
numerically, we vary $L_G$ in the simulations.

We use the same algorithm
as in \cite{spatschek07}, which even works
for dynamic elasticity.  But since we investigate the beginning of the
Grinfeld instability, the observed interface velocities are very small
in comparison to the speed of sound, and
the equations effectively reduce to the static elastic case of
Eq.~\eqref{eq:elasteqs_ph}.


While the SM model can be discretized in a relatively straightforward
manner, some care has to be taken in the other models to avoid
divisions by zero.  This is rather harmless in the GCT model, where
the problem only arises in the computation of the vectors $\normhat$
($\abs{\nabla\phi}$ goes to zero far from the interface but is
positive otherwise, so it is sufficient to ensure that the denominator
of $\normhat$ does not become smaller than a small positive number).
The main requirement in the RRV model is that $g(\phi)$ should not be
set exactly equal to zero.

In the LCT model, more attention has to be
paid to the situation where $f(\phi)$ becomes small, as will be discussed below.

Essentially, we make four types of comparison.  First, we compare the
time evolution of sinusoidal fronts (initialized with the correct
width of the profile) for a number of imposed uniaxial stresses and
obtain the linear stability spectrum numerically.  Second, we increase
the time step in the simulation for given mesh size until we reach the
maximum possible time step providing convergence to the correct
interface dynamics and then compare the achieved values.  Next, we
initialize a planar profile with the wrong interface width and observe
relaxation to a profile of the correct width.  In a realistic simulation,
slight deviations from the correct profile width may easily appear in
an initial condition for a curved interface, as analytical expressions
for constant-width profiles at arbitrary curvature are not readily
available (even for an initial germ with a shape as simple as an
ellipse it is not quite trivial to give such an expression).  Any
phase-field code should be robust against these local variations of
the profile width and should have it relax to the correct value.
Finally, we look at the evolution of an elliptical inclusion. Since
the phase-field parameter is a conserved quantity, the ellipse should
morph to a circle with the same area.

\begin{figure}[h!]
\noindent \center
\noindent
\includegraphics[width=8.5cm]{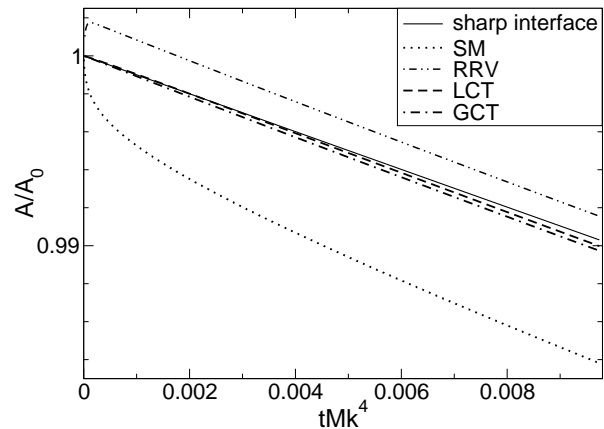}
\caption{Amplitude evolution for a uniaxial stress of $\driv = 0$, i.e., a
Griffith length $L_G=\infty$.
  \label{fig:amplit_nostress}}
\end{figure}

\begin{figure}[h!]
\noindent \center
\noindent \center
\noindent
\includegraphics[width=8.5cm]{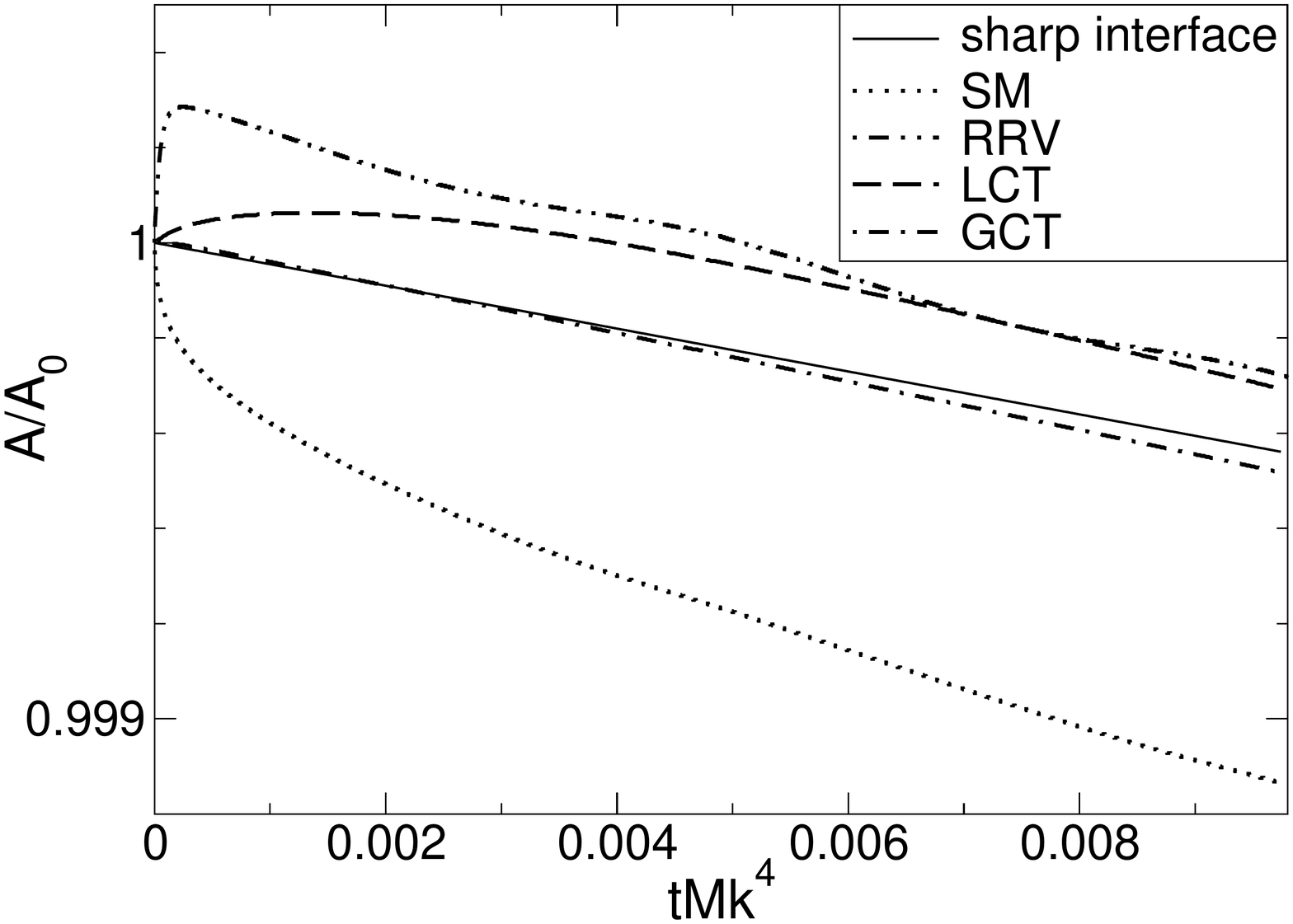}
\caption{Amplitude evolution for a uniaxial stress of $\driv = 2$.
\label{fig:amplit_smallstress}}
\end{figure}

\begin{figure}[h!]

\noindent \center
\noindent \center
\noindent
\includegraphics[width=8.5cm]{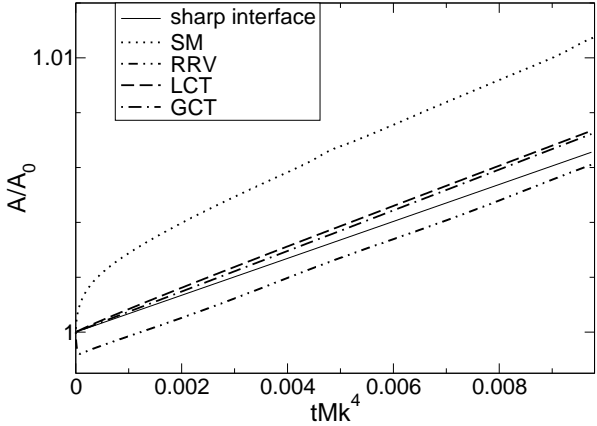}
\caption{Amplitude evolution for a uniaxial stress of $\driv = 3.5$.}
\label{fig:amplit_largestress}
\end{figure}

In Figs.~\ref{fig:amplit_nostress} to \ref{fig:amplit_largestress}, we
show the temporal evolution of a sine profile starting with a
prescribed amplitude for different values of the imposed uniaxial
stress. The four models are compared directly with the sharp interface
prediction resulting from Eq.~\eqref{eq:disprel_gf}.
Fig.~\ref{fig:amplit_nostress} exemplifies the stress-free case
discussed analytically.

All the situations considered correspond to either weak decay or weak
growth of the amplitude, as the expected exponential behavior still
appears linear on the considered time scale.

\begin{figure}[h!]
\noindent \center
\noindent
\includegraphics[width=9.0cm]{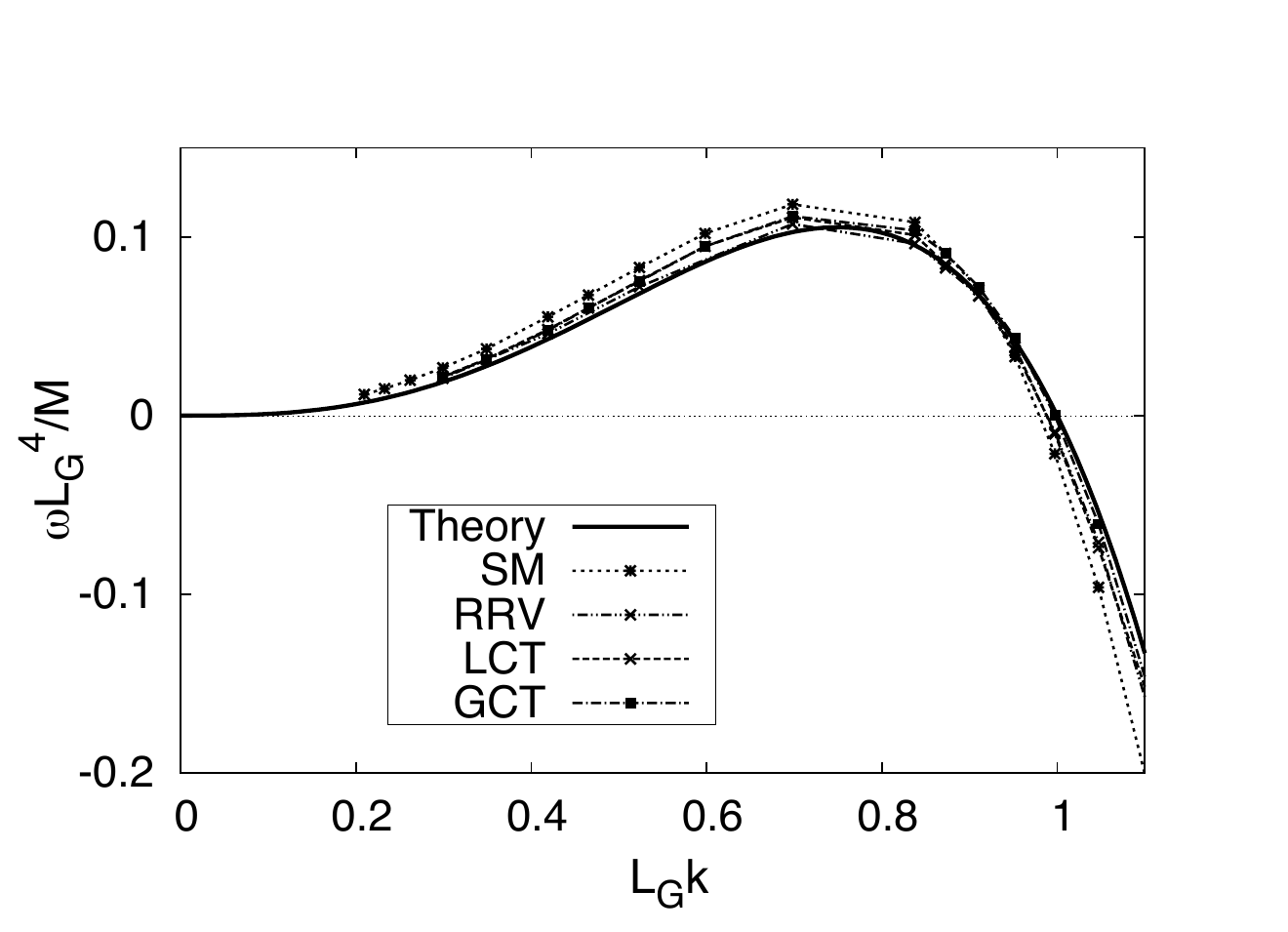}
\caption{Full spectrum of the ATG instability.}
\label{fig:fullspectrum}
\end{figure}

We note that all the models agree with the predicted behavior of the
sharp-interface limit to within better than one percent for our
parameters and time span.  While it may be observed unambiguously that
the SM model displays the largest deviation from the desired result,
one may find it surprising that it reproduces the limit so well after
all, taking into account that it does not have the right asymptotics.
Presumably the general idea mentioned after Eq.~\eqref{eq:veloceq0} is not
too far from the truth: those equations of the asymptotic behavior to
which the system can adjust locally act as an attractor for the
dynamics even before the full set of equations, implying more global
restrictions such as Eq.~\eqref{eq:veloceq0}, becomes active.  It is
striking that this seems to work even in a growth situation, where
interface velocities increase on average.

Figure \ref{fig:fullspectrum} gives a comparison of the linear
stability spectra, obtained by simulation of the four models, with the
analytical expression Eq.~\eqref{eq:disprel_gf} of the sharp-interface
model. It is pretty clear that the SM model is farthest off the
correct value both below and above the fastest-growing wavenumber.
The LCT model is good for wavenumbers above that of the
fastest-growing mode but shows stronger deviations than both the RRV
and GCT model below that mode.  The latter two models are about
equally close to the correct spectrum throughout the whole wavenumber
domain.

\begin{figure}[h!]
\noindent \center
\noindent
\includegraphics[width=8.5cm]{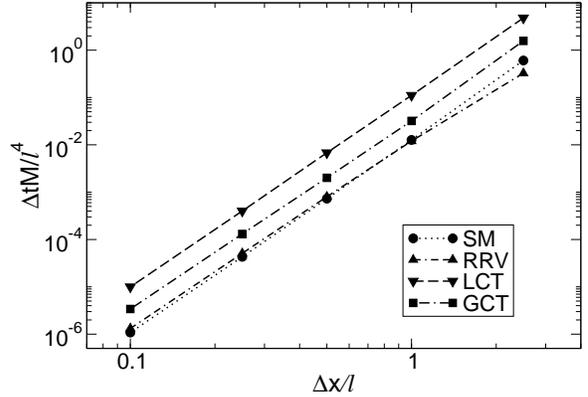}
\caption{Allowable maximal time step $\Delta t$ as a function of mesh size, where $l$ is an arbitrary length unit. Lines are a guide for the eyes only.
\label{fig:dtcomparison}}
\end{figure}

Fig. \ref{fig:dtcomparison} displays the maximum time step for a
given grid spacing leading to smooth growth where the results for all
models agreed perfectly, independent of the time-discretization.
We observe, for all models, a scaling of the maximum admissible time
step as $\Delta t\sim \Delta x^4/M$, which is not too surprising given
the fact that the equations simulated are fourth order in space and
first order in time, and we used straightforward explicit schemes for
discretization.  However, while in the two scalar models (SM, RRV)
about the same maximum time step is possible, the tensorial models allow
larger time steps; a simulation with the LCT model gains a factor of
about ten in time steps over the scalar models. While by use of
adaptive mesh techniques \cite{raetz06} (and implicit schemes), the
overall running time can certainly be reduced by more than this factor
for large systems, the advantage of the tensorial models may persist
even in such a setting as it is consistently present in a range of
grid spacings.

Next, it is interesting to compare how the different models behave
regarding their relaxation to a stationary profile when initialized
with a straight interface having a width that is either too small or
too large.  These simulations are done without elasticity, i.e., for $F=0$.
First, we verify that all the models remain in their equilibrium state
when initialized with a $\tanh$ profile of the correct width.

\begin{figure}[h!]
\noindent \center
\noindent
\includegraphics[width=8.5cm]{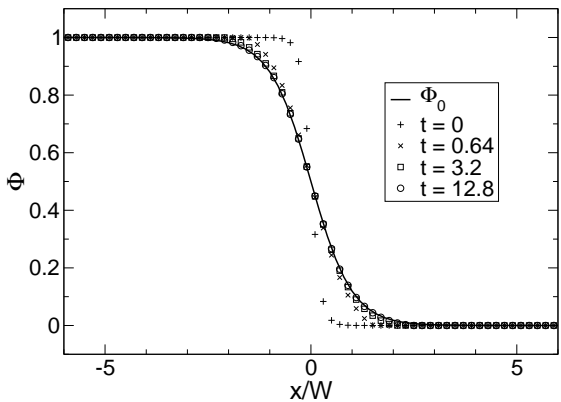}
\caption{Relaxation towards the correct interface profile for the SM model. The initial interface width is $0.25\>W$, the time $t$ is given in units of $W^4/M$.
  \label{fig:profile_SM_toothin}}
\end{figure}

For brevity of language, we define here a profile $\tanh z/W$ to have
width $W$, even when the width on which it rises from -0.9 to 0.9
rather is $2.94\>W$. In all plots, the time $t$ is given in units of
$W^4/M$, and only a small section about the interface is shown.

\begin{figure}[h!]
\noindent \center
\noindent
\includegraphics[width=8.5cm]{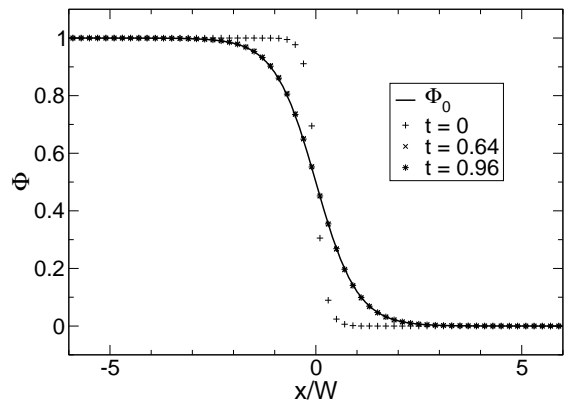}
\caption{Relaxation towards the correct interface profile for the RRV model. The initial interface width is $0.25\>W$, the time $t$ is given in units of $W^4/M$.
\label{fig:profile_RRV_toothin}}
\end{figure}

\begin{figure}[h!]
\noindent \center
\noindent
\includegraphics[width=8.5cm]{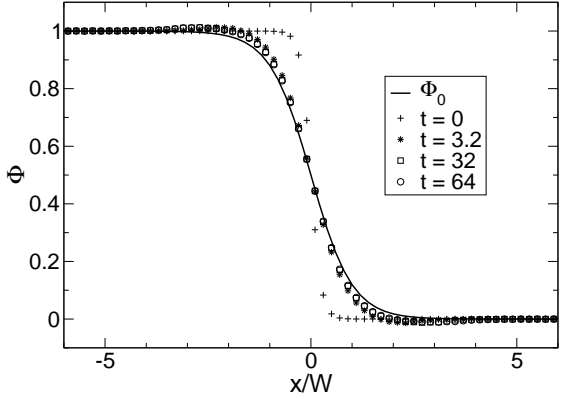}
\caption{Relaxation towards the correct interface profile for the LCT model. The initial interface width is $0.25\>W$, the time $t$ is given in units of $W^4/M$.
\label{fig:profile_LCT_toothin}}
\end{figure}

\begin{figure}[h!]
\noindent \center
\noindent
\includegraphics[width=8.5cm]{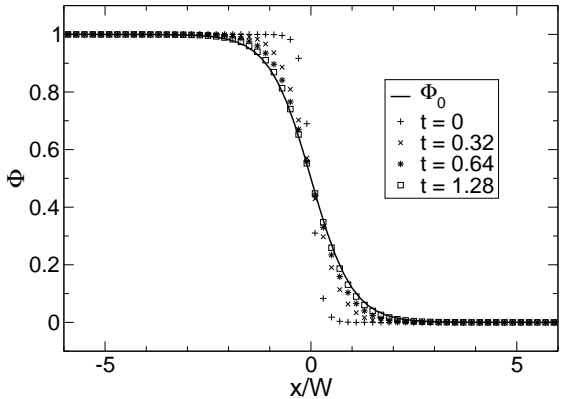}
\caption{Relaxation towards the correct interface profile for the GCT model. The initial interface width is $0.25\>W$, $N = 1.25\>M/W^2$, the time $t$ is given in units of $W^4/M$.
 \label{fig:profile_GCT_toothin}}
\end{figure}

All of the models do reasonably well in relaxing from a planar profile
of width $0.25\>W$ to their equilibrium state, see
Figs.~\ref{fig:profile_SM_toothin} to \ref{fig:profile_GCT_toothin}.
In our implementation and with the given sets of parameters,
simulations with the RRV model broke down  if the initial interface width
was chosen to be smaller than $0.23\>W$.
The RRV and GCT model relax quickly to a final
profile of width $W$, while the SM model needs a little more time (but
the permissible time \emph{step} is larger for the GCT model, so the
numerical runnig time is shortest for it).   \textcomm{cmg: For the LCT model, which
takes about as long to relax, we observe a slight overshoot of the
profile, which disappears more slowly.} In addition, some care has to
be taken in the discretization to make the LCT model deal efficiently
with too thin interfaces.

To see this, we write $Q^4 = 1-\normhat:\normhat + b_0^4
\normhat:\normhat$ with
\begin{equation}
\label{eq:defb0}
b_0 =1- \frac{W^2 (\nabla\phi)^2}{4 f(\phi)}\>.
\end{equation}
Taking an interface of width $\xi$ with the profile
$\phi(z) = \frac12 \left(1-\tanh {z}/{\xi}\right)$
we find
\begin{equation}
\label{eq:resb0}
b_0^4 =\left(1-\frac{W^2}{\xi^2}\right)^4\>,
\end{equation}
which becomes very large for $\xi\ll W$.  In fact, for $\xi=0.25 W$, we
have $b_0^4=50625$, a number that, when plugged into the equations of
motion, would impose a prohibitively small time step for stability (or
accuracy, in an implicit scheme).  Hence, we introduce a cutoff for
$b_0^4$ on the order of 50. In production runs, where one normally
starts with a front profile having at least approximately the correct
width, a cutoff of 10 may be sufficient.

\begin{figure}[h!]
\noindent \center
\noindent
\includegraphics[width=8.5cm]{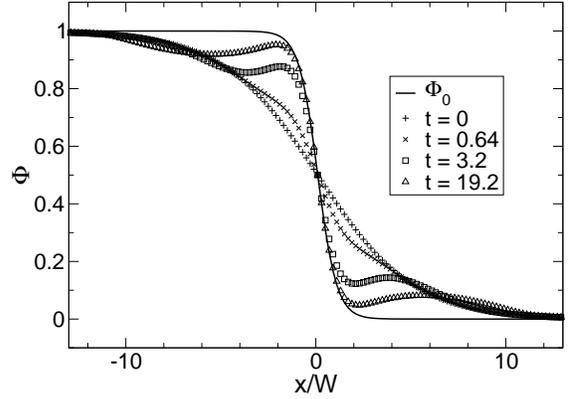}
\caption{Relaxation towards the correct interface profile for the SM model. The initial interface width is $5\>W$, the time $t$ is given in units of $W^4/M$.
\label{fig:profile_SM_toothick}}
\end{figure}

\begin{figure}[h!]
\noindent \center
\noindent
\includegraphics[width=8.5cm]{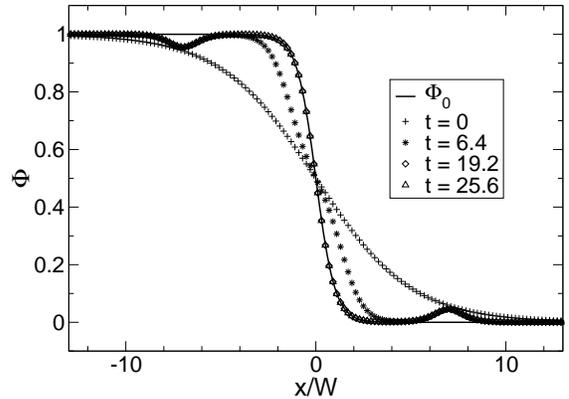}
\caption{Relaxation towards the correct interface profile for the RRV model. The initial interface width is $5\>W$, the time $t$ is given in units of $W^4/M$.
\label{fig:profile_RRV_toothick}}
\end{figure}

\begin{figure}[h!]
\noindent \center
\noindent
\includegraphics[width=8.5cm]{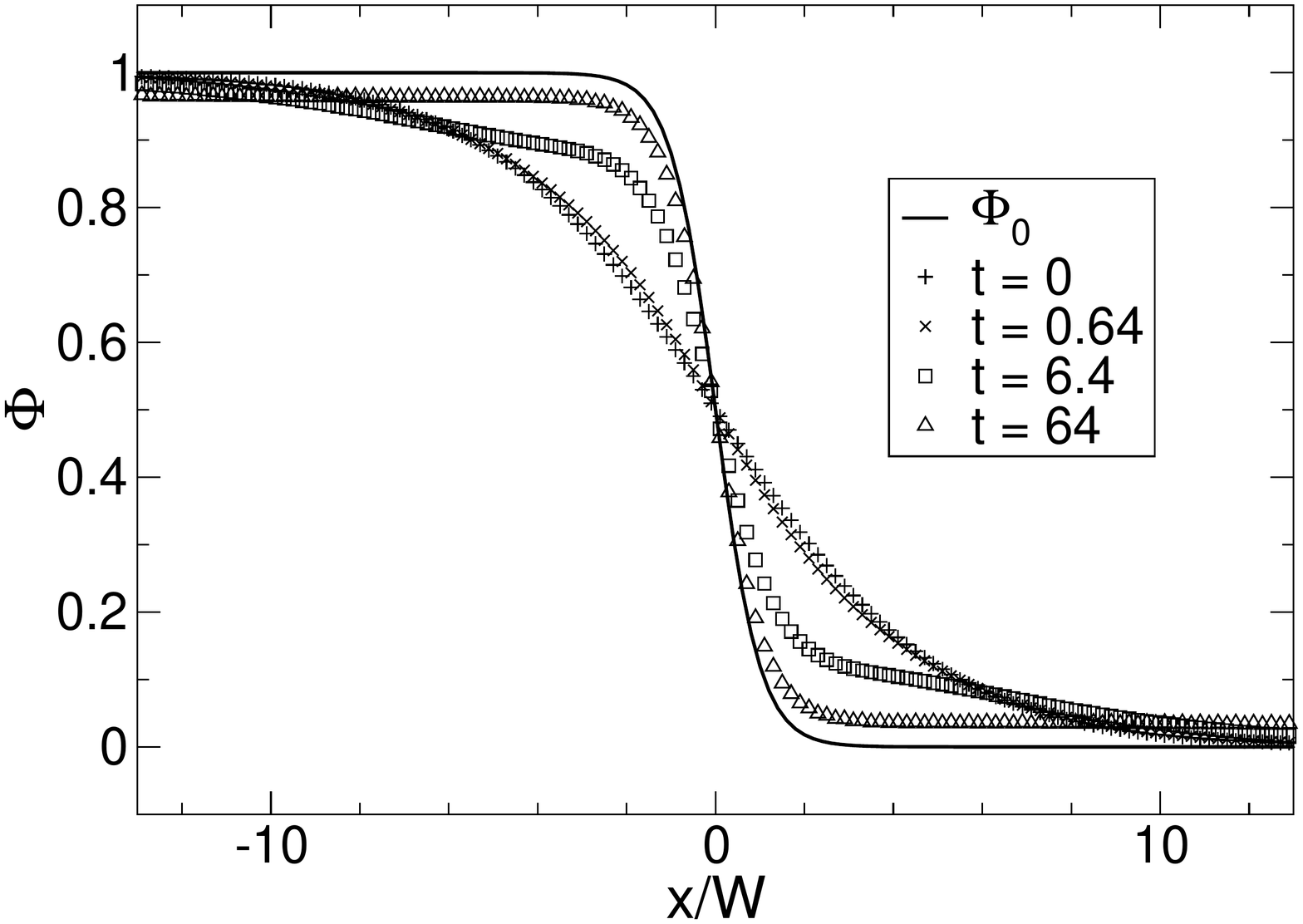}
\caption{Relaxation towards the correct interface profile for the LCT model.
The initial interface width is  $5\>W$, the time $t$ is given in units of $W^4/M$. \label{fig:profile_LCT_toothick}}
\end{figure}

\begin{figure}[h!]
\noindent \center
\noindent
\includegraphics[width=8.5cm]{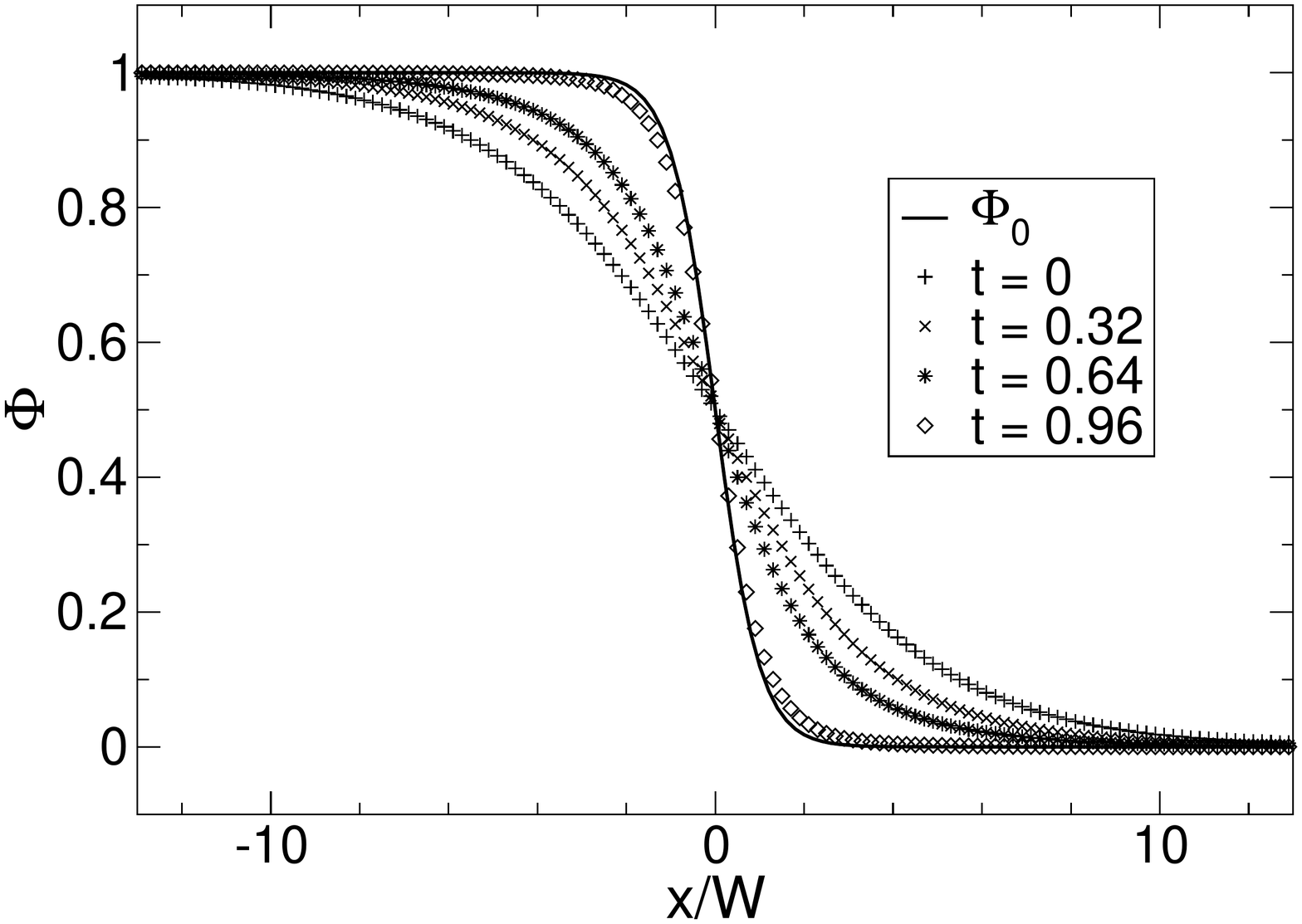}
\caption{Relaxation towards the correct interface profile for the GCT model.
The initial interface width is $5\>W$, the time $t$ is given in units of $W^4/M$. \label{fig:profile_GCT_toothick}}
\end{figure}

When the profile is initialized with too large a width
[Figs.~\ref{fig:profile_SM_toothick} through
\ref{fig:profile_GCT_toothick}], more interesting differences can be
seen.  Not unexpectedly, the GCT model
[Fig.~\ref{fig:profile_GCT_toothick}] is the one making the least fuss
about an interface five times too wide: that the model is
nonconservative on the scale where the phase field varies strongly is
an advantage here.
The interface approaches its correct width in a time of about $t \approx 1.25\>W^2/N$, which corresponds to $t\approx 1\>W^4/M$ for our parameter choice.
For the other models, this takes much longer, as this
kind of adaptation requires diffusion orthogonally to the front, which
is slow because it is suppressed in the asymptotic limit.

The RRV model goes through a series of transformations of the profile
involving as an intermediate state a spatially varying slope in the
vicinity of the contour line $\phi=\frac12$ defining the interface
position.  Even after a time of $t \approx 30\>W^4/M$, while near the interface
position the profile is well-behaved and has the right width, there
are still indentations in it far from the interface, and these
disappear only slowly.

While the LCT model keeps a nicer profile all the time, it relaxes
only slowly as well.  Moreover, if the boundary values of $\phi$ are
not fixed to be equal to zero or one, it will relax to constant values
in the bulk different from these ideal values (in the absence of
elasticity).  Indeed, inspection of Eq.~\eqref{eq:phidyn_cval} shows
immediately that any constant value of $\phi$ solves the bulk
equations of motion.  (This is no longer true in the presence of
elasticity.)  For phases extending to the system boundary, the value
of the constant is only fixed by the boundary conditions.  Therefore,
the model should always be run with Dirichlet boundary conditions for
the phase field.  (Due to the conservation law, inclusions of one
phase in another will keep their $\phi$ value, even in the presence of
elasticity, if correctly initialized to zero or one, as long as their
inner volume is much larger than that of their interface.)  Performing
such a simulation, we found relaxation to be as slow as for the SM
and RRV models but the interface profile to look more reasonable.

To summarize, when interface thickness is believed to be an issue in
simulations, i.e., when there are reasons to think that it might vary
considerably (which may be the case when surface tension anisotropy is
included in the model), the nonexact realization of the conservation
law by the GCT model may turn out a virtue rather than a drawback,
since changes in the direction normal to the interface by diffusion
only, as realized in the other models, tend to be too slow.

\begin{figure}[h!]
\noindent \center
\noindent
\includegraphics[width=8.5cm]{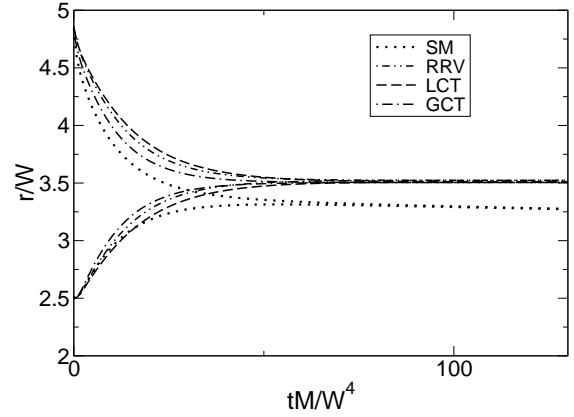}
\caption{Comparison of the time development of the size of an elliptical
  inclusion. The square system had the length $L/W = 20$.  The
  initial ellipse had a semimajor of $a_0=L/4$ and a semiminor of $L/8$.
  All models except the SM-model converge to circles with the
  same radius $r$.
  \label{fig:ellipse_comparison}}
\end{figure}

Finally, we compare the different dynamics for a ''real-life" situation
of an elliptical inclusion that morphs into a circle without elastic effects, $F=0$. The system is
initialized with a sharp interface ellipse with semimajor $a_0$ and
semiminor $a_0/2$ and is then allowed to relax for a few time steps
running the GCT-model (with a Lagrange multiplier $\la = 0$), in order
to obtain an initial condition with the correct interface width
everywhere. We then measure the time evolution of the semimajor and
semiminor of the ellipse, continuing the run with the model to be
studied.

\begin{figure}[h!]
\noindent \center
\noindent
\includegraphics[width=8.0cm]{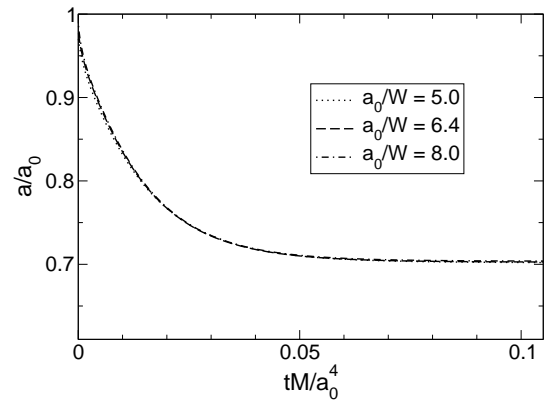}
\caption{Elliptical inclusion: comparison of the time development of the length
of the semimajor $a$ for the LCT-model. The initial length is denoted by $a_0$ and the different curves
correspond to different scale separations $a_0/W$. All curves collapse onto a single line.
  \label{fig:ellipse_LCT_scaled}}
\end{figure}

\begin{figure}[h!]
\noindent \center
\noindent
\includegraphics[width=8.0cm]{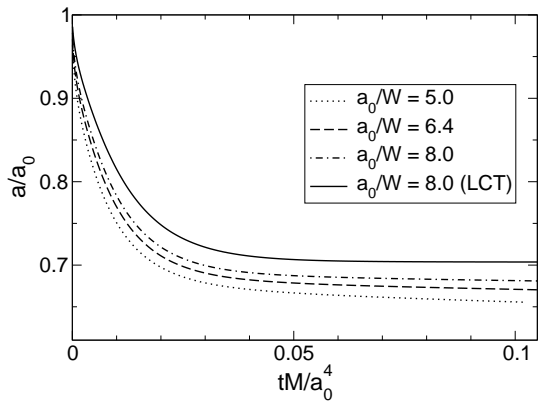}
\caption{Elliptical inclusion: comparison of the time development of the length
of the semimajor $a$. The initial length is denoted by $a_0$ and the different curves
correspond to different scale separations $a_0/W$. The performance of the
SM-model becomes asymptotically better for larger
  systems. Results for the LCT-model have been included as a
  reference.
  \label{fig:ellipse_comparison_scaled}}
\end{figure}

As Fig.  \ref{fig:ellipse_comparison} shows, all models but the
SM-model converge to a circle with the correct radius $\sqrt{2}
a_0/2$.  The SM-model shows different behavior, namely a too
small radius that seems to decrease further.  Since the phase field is a
conserved quantity (we also checked that numerically for our code),
this can only mean that the final shape of the inclusion is not a true
circle but a slightly deformed one, displaying a certain level of
anisotropy.  We then increase the size of the system and the included
ellipse while keeping the interface width constant, resulting in a
better scale separation $a_0/W$.  While for the LCT-, GCT- and RRV-
model the curves collapse onto a single line, this is not the case for
the SM-model. Fig.~\ref{fig:ellipse_LCT_scaled} demonstrates this behavior for the LCT-model.  The comparison for the SM-model is shown in Fig.~\ref{fig:ellipse_comparison_scaled}.

\section{Conclusions\label{sec:conclusions}}

The intuitive approach to constructing a phase-field model for surface
diffusion consists in using the chemical potential known from the
nonconservative model to define a current, involving its gradient and
a mobility that vanishes in the bulk phases, and in taking the
divergence of this current as the time derivative of the phase field.
As has been shown in this article, this approach -- the SM model --
fails to produce the correct asymptotics in a subtle way.  It does
reproduce the equlibrium limit correctly and it appears to work
numerically, although less efficiently than the alternatives
discussed.

We offer a simple argument why the SM model should not be expected to
work properly: The chemical potential functional of the model is
constructed so that the chemical potential vanishes in the bulk
phases.  As the diffusion operator is essentially a scalar, diffusion
acts also orthogonally to the interface; in its vicinity, the effect
is even strong, because the slope of the phase field is largest in the
direction perpendicular to the corresponding level set.  This
diffusive effect constitutes a driving force for relaxation of the
chemical potential towards zero also close to the interface
(asymptotically, the chemical potential \emph{is} zero at next-to
leading order).  Surface diffusion of the chemical potential is then
not the only effect contributing to the interface dynamics.

The RRV model avoids this problem by leaving the chemical potential in
the bulk undetermined.  Absence of diffusion in the bulk is not
guaranteed by the chemical potential but by the vanishing mobility.
Since the bulk chemical potential is free to vary, a true interface
chemical potential can build up, the surface diffusion of which
governs the interface dynamics.

Our contribution in this article is to explore the idea that the
failure of the SM model might be remedied instead by making the
mobility a tensor.  After all, surface diffusion may be interpreted as
highly anisotropic three-dimensional diffusion with a diffusion tensor
that has zero eigenvalue in one direction. Whereas the preexisting RRV
model has the correct asymptotics, it does not exploit that idea.  A
straightforward attempt of its realization however fails in a rather
drastic way, because restricting diffusion to the surfaces of constant
phase field does not impose any functional dependence of $\phi$ in the
normal direction given by this foliation.

Modifying the tensorial mobility, one obtains the LCT and GCT models,
both exhibiting the correct asymptotic behavior.

An analytic linear stability analysis of a planar front demonstrates
that these two models reproduce the correct dispersion relation for
local perturbations of the profile corresponding to a modified
interface position. The phase-field dispersion relation
is even free of corrections due to the
front width. A similar analysis turns out infeasible for the SM model,
while it could in principle be completed numerically for the RRV model
(after analytic reduction to an ordinary differential equation).

Numerical study of the four models suggests that whereas the SM model
has a range of quantitative validity (despite its not being
asymptotic), and the RRV model is definitely viable, the modified
tensorial models discussed here are probably more useful for
large-scale simulations, as they permit larger time steps.

A possible way to understand this is as follows.  In the
scalar-mobility models (SM and RRV), the inner equation determining
the interface velocity appears only three orders in $W$ after the
leading order.  A simulation must of course represent the full model
equations.  Suppose we wish the interface velocity to be determined
with an error not exceeding order $W$.  Then the leading-order
equation must be simulated with an accuracy $O(W^4)$ at least.  If one
takes a simple second-order accurate discretization for the gradient
operators in the equations, the numerical error due to discretization
alone will be $O(\Delta x^2)$ for a grid spacing $\Delta x$.  To keep
this smaller than $W^4$, $\Delta x$ would have to scale with
$W^2$, which can be expected to lead to large computation times.
Therefore, in the scalar mobility models, high-accuracy
discretizations are mandatory, even when the asymptotic error is
controllable, as is the case for RRV.

On the other hand, in the LCT and GCT models, discussed in
Secs.~\ref{sec:fct} and \ref{sec:act}, as soon as the phase-field
profile is represented with an error of order $W$ or better, the
effective leading order is only one order lower than the one
determining the interface velocity, similar to the nonconservative
case.  Hence, reasonable accuracy should be attainable with grid
spacings that scale as $W$, not as $W^2$.  None of the benefits of
high-accuracy discretizations would be lost to the need of
representing terms very accurately that are very small in the
leading-order equation.

Generally speaking, the most efficient model in terms of computational
cost is the LCT model.  However, it is slowest in terms of ``real''
time (though not in terms of the number of time steps), when it comes
to the relaxation from a wrong width of a planar interface to the
correct value.  This may be seen as a signature that the model is most
efficient in suppressing diffusion perpendicular to the interface.  In
optimally initialized simulations, interface width variations arise
gradually, via curvature changes and/or orientation changes (in models
with anisotropic surface tension) and one should expect their
relaxation via diffusion \emph{along} the interface to be sufficiently
efficient.  Nevertheless, this property of the LCT model reduces its
robustness as a numerical tool.

This is why we suggest the GCT model as an alternative that seems to
be more accurate in most applications than the LCT model and digests
variations of the interface width more easily than all the other
models. Both accuracy and robustness of the GCT model, connected with
its still favorable efficiency and simplicity of implementation,
should make it the approach of choice in most cases.

\textbf{Acknowledgments} We wish to thank E. Brener and H.
M\"uller-Krumbhaar for stimulating discussions that have led, among
other things, to the introduction of a Lagrange parameter into the GCT
model.  Support of this work by DFG grants KA 672/9-1 and \hbox{MU 1170/6-1},
and the U. S. Department of Energy, Office of Basic Energy Sciences,
through the Computational Materials Science Network program is gratefully acknowledged.


\appendix

\section{Matching conditions\label{sec:matching}}

Let $\tilde\psi(x,z,t)=\psi(r,\s,t)$ be some arbitrary (sufficiently
often differentiable) function of space and time obtained in solving
the outer equations.  We write the corresponding function of the inner
solution as $\Psi(\rho,\s,t)$, and suppress from now on, in this
section, the dependence of functions on $\s$ and $t$.  Moreover, we
write the coefficient functions in expansions with respect to $W$ with
simple subscripts indicating their order rather than superscripts in
parentheses as in the main text.  There, the notation is dictated by
the fact that a subscript would interfere with certain other
subscripts; here, a superscript would interfere with the primes
denoting derivatives.

We must have the asymptotic relationship
\begin{equation}
  \label{eq:asym_match1}
  \Psi(\rho) \sim \psi(r) = \psi(W\rho)
\quad (\rho\to\infty,\,W\to0,\,W\rho\to0)\,.
\end{equation}
Expanding both functions in powers of $W$, we get
\begin{eqnarray}
  \Psi(\rho) &=& \Psi_0(\rho)+W\Psi_1(\rho)
                  +W^2\Psi_2(\rho)+\dots \>,    \label{eq:psi_exps1}\\
   \psi(W\rho) &=& \psi_0(r)+W\psi_1(r)
                  +W^2\psi_2(r)+\dots   \nonumber\\
               &=&  \psi_0(0)+W[\rho\psi'_0(0)+\psi_1(0)] \nonumber\\
                &&\mbox{}  +W^2[\rho^2\frac12\psi''_0(0)
         +\rho\psi'_1(0)+\psi_2(0)]+\dots \>, \nonumber\\
 \hspace*{1cm}\label{eq:psi_exps2}
\end{eqnarray}
where the derivatives are to be taken for $r\to +0$, should they be
discontinuous at $r=0$.  Analogous expressions with $r\to -0$ are
obtained for the asymptotics as $\rho\to-\infty$.

Equating powers of $W$, we then successively get the asymptotic
relationships
\begin{eqnarray}
 \lim_{\rho\to\pm\infty} \Psi_0(\rho) &=& \psi_0(\pm0)  \>, \label{eq:asymrels1}\\
 \Psi_1(\rho)  &\sim& \rho\psi'_0(\pm0)+\psi_1(\pm0) \quad (\rho\to\pm\infty)
\>,\nonumber\\ \label{eq:asymrels2}\\
 \Psi_2(\rho)  &\sim& \frac12\rho^2\psi''_0(\pm0)+\rho\psi'_1(\pm0)
  +\psi_2(\pm0) \nonumber\\
                  && \quad\quad\quad\quad\quad\quad    (\rho\to\pm\infty) \>.
\label{eq:asymrels3}
\end{eqnarray}
Moreover, asymptotic relations such as \eqref{eq:asymrels2} can be
decomposed into statements about function limits
\begin{eqnarray}
\lim_{\rho\to\pm\infty} \partial_\rho\Psi_1(\rho) &=& \psi'_0(\pm0) \>,
 \label{eq:asymrels2a} \\
\lim_{\rho\to\pm\infty} \left[\Psi_1(\rho)- \rho\psi'_0(\pm0)\right]
 &=& \psi_1(\pm0) \>.
 \label{eq:asymrels2b}
\end{eqnarray}

\section{Useful properties of the phase field functions\label{sec:collection}}

In order to simplify it for the reader to find the actual
relationships for the various functions involving the phase-field that
are used in the text, they are collected here for reference (and
concreteness). Often only certain properties but not the precise form
of these functions are important.

The chosen double-well potential is
\begin{equation}\label{eq:fphi}
    f(\phi) = \phi^2 (1-\phi)^2 \>.
\end{equation}
Its derivative is given by
\begin{equation}\label{eq:fprimephi}
    f'(\phi) = 2 \phi (1-\phi) (1-2\phi) \>,
\end{equation}
its second derivative reads
\begin{equation}\label{eq:fprimeprimephi}
    f''(\phi) = 2 \left[1-6 \phi(1-\phi)\right] \>.
\end{equation}
Let us define another function $h(\phi)$, which turns out useful, by
\begin{equation}\label{eq:hphi}
    h(\phi) = \phi^2 (3-2\phi) \>,
\end{equation}
having the derivative
\begin{equation}\label{eq:hprimephi}
    h'(\phi) = 6 \phi (1-\phi)  \>.
\end{equation}


To solve the ordinary differential equation satisfied by the
zeroth-order inner solution
\begin{equation}\label{eq:diffeq_Phi0}
    \partial_{\rho\rho} \Phinul -2 f'(\Phinul) = 0 \>,
\end{equation}
with boundary conditions $\lim_{\rho\to-\infty}\Phinul(\rho) = 1$ and
$\lim_{\rho\to\infty}\Phinul(\rho) = 0$, we multiply by $\partial_\rho
\Phinul$, integrate and take the square root (with the correct sign) to
obtain
\begin{equation}\label{eq:diffeq_Phi0_firstint}
    \partial_\rho \Phinul = -2 \Phinul (1-\Phinul)
= -\frac13 h'(\Phinul)
    \>,
\end{equation}
which can be solved by separation of variables. The solution is, up to
a translation in $\rho$, given by
\begin{equation}\label{eq:solution_Phi0}
    \Phinul = \frac12 (1-\tanh\rho) \>.
\end{equation}
Requiring the position of the interface to be at $\rho=0$ fixes the
choice out of the one-parameter set of solutions, present due to the
translational invariance of the differential equation.

With the help of the second equality of
\eqref{eq:diffeq_Phi0_firstint}, it is easy to calculate certain
integrals appearing in the asymptotic analysis.  Those integrals
typically contain the factor $\left(\partial_\rho \Phinul\right)^2$;
to do the integral, it is then beneficial to replace one of the
factors (and only one) with $-h'(\Phinul)/3$.  Integrals obtained this
way have the structure
\begin{align}
 \label{eq:typicalints}
I = &\int_{-\infty}^{\infty}\,d\rho\> f(h(\Phinul))\,
\,\left(\partial_\rho \Phinul\right)^2
 \nonumber\\
 = & -\frac13 \int_{-\infty}^{\infty}\,d\rho\> f(h(\Phinul))\, h'(\Phinul)
\,\partial_\rho \Phinul
\nonumber\\
 = & -\frac13\int_1^0 \,d\Phinul \>  f(h(\Phinul)) h'(\Phinul) \nonumber\\
= & \frac13\int_0^1
\,d\hnul\> f(\hnul)
\>.
\end{align}
This way, one arrives, for example, at
\begin{equation}
  \label{eq:intdphirho2}
  \int_{-\infty}^{\infty} \left(\partial_\rho \Phinul\right)^2
 d\rho = \frac13 \>.
\end{equation}

In the RRV model \cite{raetz06}, we have to evaluate two more integrals, namely
\begin{equation}
  \label{eq:intaltrho}
 \begin{aligned}
 \int_{-\infty}^{\infty} g(\Phinul)  \,\partial_\rho \Phinul  d\rho
& = -\int_0^1 10 x^2 (1-x)^2 dx = -1/3\>,\\
   \int_{-\infty}^{\infty} B(\Phinul)\, d\rho & =
 \int_{-\infty}^{\infty} 3 \left(\partial_\rho \Phinul\right)^2 =1 \>.
\end{aligned}
\end{equation}

\section{Details of the analytic linear stability
 calculation\label{sec:linstab_det}}

In the nonconservative case, the eigenvalue problem to be solved reads
\begin{equation}
\label{eq:eigenval_nc_inn}
\omega\Psi = \frac{M}{W^2}\left(\partial_{ZZ}-W^2 k^2
 -2 f''(\Phizero)\right) \Psi\>.
\end{equation}
This may be considered the inner problem, given in the coordinates
($x$,$Z$), with $Z=z/W$. Setting $\psi(z) = \Psi(Z)$, the
corresponding outer problem reads:
\begin{equation}
\label{eq:eigenval_nc_out}
\omega\psi = M\left(\partial_{zz}- k^2
 -\frac{4}{W^2} \right) \psi\>,
\end{equation}
where we have used $\lim_{Z\to\pm\infty} f''(\Phizero)=2$ [see
\eqref{eq:fprimeprimephi}].  Using the expansions
\begin{align}
\label{eq:linstab_expansion}
\omega &= \frac{\omega_{-2}}{W^2}+ \frac{\omega_{-1}}{W} +
\omega_{0} + \omega_{1} W + \dots\nonumber\\
\Psi(Z) &= \Psi_0(Z) + W \Psi_1(Z) + W^2 \Psi_2(Z) + \dots\\
\psi(z) &= \psi_0(z) + W \psi_1(z) + W^2 \psi_2(z) + \dots\nonumber
\end{align}
we then have, at leading order
\begin{equation}
\omega_{-2}\psi_0 = -4M \psi_0
\end{equation}
meaning that either $\omega_{-2}=-4M$ or $\psi_0\equiv 0$. Thus we
have to distinguish two cases.

If $\omega_{-2}=-4M$, the lowest-order inner equation becomes
\begin{equation}
\label{eq:schroed_1d_a}
\Psi_0''(Z)+ \frac{6}{\cosh^2 Z} \Psi_0(Z) =0\>,
\end{equation}
where we have used \eqref{eq:solution_Phi0}.  This is a
one-dimensional Schr\"o\-din\-ger equation that can be reduced to Legendre's
differential equation \cite{abramowitz72} via the substitution
$u=\tanh Z$ and hence is exactly solvable. In this particular case,
the solution that remains finite at infinity is the second-order (in $u$) Legendre
polynomial
\begin{equation}
\label{eq:solschr_1}
\Psi_0(Z) = \frac12 \left(3\tanh^2 Z-1 \right)\>.
\end{equation}
This means that the perturbation is not localized -- it does not
approach zero for $Z \to\pm\infty$.  As usual, once the zeroth-order
solution has been found, perturbation theory can be carried through
without major difficulties, inserting the expansions
\eqref{eq:linstab_expansion} into \eqref{eq:eigenval_nc_inn} and
\eqref{eq:eigenval_nc_out}. Calculating the eigenvalue to zeroth order
in $W$, we find $\omega_a$ from \eqref{eq:spectr_nc}. In fact,
continuing the calculation to higher orders, we note that no
additional contributions to $\omega_a$ arise.  The suspicion that
\eqref{eq:solschr_1} is an exact solution -- and $\omega_a$ the
corresponding exact eigenvalue -- to Eq.~\eqref{eq:eigenval_nc_inn} is
confirmed by backsubstitution into the equation.

The second solution is obtained by assuming $\omega_{-2}=0$ and,
hence, $\psi_0(z)=0$ in the outer region.  This gives the leading-order inner
equation
\begin{equation}
\label{eq:schroed_1d_b}
\Psi_0''(Z)- 2 f''(\Phizero) \Psi_0(Z) =0\>,
\end{equation}
the relevant solution of which we know already, because the linear
operator acting here is just $\linopl$ from \eqref{eq:linop} with the
role of $\rho$ taken by $Z$.  Hence
\begin{equation}
\label{eq:solschr_2}
\Psi_0(Z) = \Phizero'(Z) = -\frac{1}{2\cosh^2 Z}\>,
\end{equation}
which is a solution localized about the interface and approaching zero
for $Z \to\pm\infty$. Inserting the expansions
\eqref{eq:linstab_expansion} with the evaluated results for
$\omega_{-2}$, $\psi_0(z)$ and $\Psi_0(Z)$ into
\eqref{eq:eigenval_nc_inn} and \eqref{eq:eigenval_nc_out}, we find
that the eigenvalue is to all orders given by $\omega_b$ from
\eqref{eq:spectr_nc} and that, in fact, $\Psi_0(Z)$ and $\omega_b$
constitute an exact solution to the eigenvalue problem. There may be
more admissible solutions, but for all of them $\omega$ should be
negative and diverge more strongly than $1/W^2$ for $W\to 0$, so they
will not be relevant in the limit of small $W$.


\begin{thebibliography}{22}
\bibitem{osher88} S.  Osher and J. A. Sethian
J. Comp. Phys., {\bf 79}, 12 
(1988).

\bibitem{sethian99} J. A. Sethian, \emph{Level Set Methods and Fast Marching
  Methods: Evolving Interfaces in Computational Geometry, Fluid
  Mechanics, Computer Vision, and Materials Science} (2nd ed.),
  Cambridge University Press (1999).

\bibitem{karma96}
A. Karma and W.-J. Rappel, Phys. Rev. E {\bf 53},  R3017  (1996).

\bibitem{karma98} A. Karma and W.-J. Rappel, Phys. Rev. E {\bf 57},
  4323 (1998).


\bibitem{langer75}
J.~S. Langer and R.~F. Sekerka, Acta Metall. {\bf 23}, 1225 (1975).

\bibitem{collins85}
J. B. Collins and H. Levine, Phys. Rev. B {\bf 31}, 6119 (1985).

\bibitem{caginalp86}
G. Caginalp and P. Fife,  Phys. Rev. B {\bf 33}, 7792 (1986).

 \bibitem{mcfadden93}
G.B. McFadden, A.A. Wheeler, R.J. Braun, and S.R. Coriell,
Phys. Rev. E {\bf 48}, 2016 (1993).


\bibitem{kobayashi93} R. Kobayashi, Physica D {\bf 63}, 410 (1993).

\bibitem{kobayashi94} R. Kobayashi, Exp. Math. {\bf 3},  60  (1994).


\bibitem{abel97} T. Abel, E. Brener, and H. M{\"{u}}ller-Krumbhaar,
  Phys. Rev. E {\bf 55}, 7789 (1997).

\bibitem{wheeler96} A.A. Wheeler, G.B. McFadden, and W.J. Boettinger,
Proc. Roy. Soc. Lond. A {\bf 452}, 495 (1996).

\bibitem{garcke99}
H. Garcke, B. Nestler, and B. Stoth,
SIAM J. Appl. Math. {\bf 60}, 295 (1999).

\bibitem{opl03} O. Pierre-Louis, 
Phys. Rev. E {\bf 68}, 021604 (2003).

\bibitem{voigt04} F.Otto, P. Penzler, A. R\"atz, T. Rump, A. Voigt,
 Nonlinearity {\bf 17}, 477 (2004).

\bibitem{mueller99}
J. M\"uller and M. Grant, Phys. Rev. Lett. {\bf 82},  1736  (1999).

\bibitem{kassner99}
K. Kassner and C. Misbah, Europhys. Lett. {\bf 46},  217  (1999).

\bibitem{kassner01}
K. Kassner, C. Misbah, J. M\"uller, J. Kappey, P. Kohlert,
Phys. Rev. E {\bf 63}, 036117  (2001).

\bibitem{grinfeld86}
M. Grinfeld, Sov. Phys. Dokl. {\bf 31},  831  (1986).

\bibitem{asaro72}
R. Asaro and W. Tiller, Metall. Trans. {\bf 3},  1789  (1972).

\bibitem{yang93}
W.~H. Yang and D.~J. Srolovitz, Phys. Rev. Lett. {\bf 71}, 1593 (1993).

\bibitem{kassner94}
K. Kassner and C. Misbah, Europhys. Lett. {\bf 28},  245  (1994).

\bibitem{haataja02} M. Haataja, J. M\"uller, A.D. Rutenberg, M.
  Grant,
  Phys. Rev. B {\bf 65}, 165414 (2002).

\bibitem{raetz06} A. R\"atz, A. Ribalta, A. Voigt,
J. Comp. Phys. {\bf 214}, 187 
(2006)

\bibitem{yeon06} D.-H. Yeon, P.-R. Cha, M. Grant,
Acta Materialia {\bf 54},  1623 
(2006)

\bibitem{cahn96} J.W. Cahn, C.M. Elliott, A. Novick-Cohen,
Euro. J. Appl. Math. {\bf 7}, 287 (1996).

\bibitem{mahadevan99a}
M. Mahadevan and R.M. Bradley,
Physica D {\bf 126}, 201 (1999).


\bibitem{bhate00}
D.N. Bhate, A. Kumar, and A.F. Bower,
J. Appl. Phys. {\bf 87}, 1712 (2000).

\bibitem{kassner_condmat06}
K. Kassner, {\em How to model surface diffusion using the phase-field approach},
cond-mat/0607823



\bibitem{spivak79}
M.A. Spivak,  Comprehensive Introduction to Differential Geometry, Vol.
3, 2nd ed. 
Publish or Perish, Inc., Wilmington, Del., 1979.

\bibitem{spatschek07}
R. Spatschek, C. M\"uller-Gugenberger, E. Brener, and B. Nestler,
 Phys. Rev. E {\bf 75}, 066111 (2007).

\bibitem{abramowitz72} M. Abramowitz, I. Stegun, \emph{Handbook of
    Mathematical Functions}, Dover, New York (1972).


\end{thebibliography}
\end{document}